\documentclass[namedreferences,hyperref,optionalrh]{spr-sola}
\usepackage{graphicx}        
\usepackage{color}           

\chardef\us=`\_

\begin{document}

\begin{frontmatter}
\title{The Wide Field Imager (WFI) Instruments for the Polarimeter to Unify the Corona and Heliosphere (PUNCH)}

\author[addressref={aff-swri},corref,email={glenn.laurent@swri.org}]
   {\inits{G.T.}\fnm{Glenn~T.}\ \lnm{Laurent}\orcid{0009-0000-6578-326X}}
\author[addressref={aff-swri},email={craig.deforest@swri.org}]
   {\inits{C.E.}\fnm{Craig~E.}\ \lnm{DeForest}\orcid{0000-0002-7164-2786}}
\author[addressref={aff-swri}]
   {\inits{M.N.}\fnm{Matt~N.}\ \lnm{Beasley}\orcid{0009-0004-0646-0523}}
\author[addressref={aff-swri}]
   {\inits{N.F.}\fnm{Nicholas~F.}\ \lnm{Erickson}\orcid{0000-0001-6028-1703}}
\author[addressref={aff-swri2}]
   {\inits{R.R.}\fnm{Roy~R.}\ \lnm{Graham}}
\author[addressref={aff-swri}]
   {\inits{M.H.}\fnm{Mary~H.}\ \lnm{Hanson}\orcid{0009-0005-8304-4037}}
\author[addressref={aff-swri}]
   {\inits{J.M.}\fnm{J.~Marcus}\ \lnm{Hughes}\orcid{0000-0003-3410-7650}}
\author[addressref={aff-swri}]
   {\inits{D.A.}\fnm{Derek~A.}\ \lnm{Lamb}\orcid{0000-0002-6061-6443}}
\author[addressref={aff-swri2}]
   {\inits{R.}\fnm{Reith}\ \lnm{Nolan}}
\author[addressref={aff-swri}]
   {\inits{S.}\fnm{Steve}\ \lnm{Osterman}\orcid{0009-0000-0848-5805}}
\author[addressref={aff-swri2}]
   {\inits{T.}\fnm{Trent}\ \lnm{Peterson}}
\author[addressref={aff-swri}]
   {\inits{M.}\fnm{Michael}\ \lnm{Shoffner}}
\author[addressref={aff-swri2}]
   {\inits{K.D.}\fnm{Kelly~D.}\ \lnm{Smith}}
\author[addressref={aff-swri}]
   {\inits{T.}\fnm{Travis}\ \lnm{Smith}}
\author[addressref={aff-swri2}]
   {\inits{T.}\fnm{Todd}\ \lnm{Veach}\orcid{0000-0003-2347-6725}}
\author[addressref={aff-swri2}]
   {\inits{W.L.}\fnm{William~L.}\ \lnm{Wells}}
\author[addressref={aff-swri2}]
   {\inits{A.J.}\fnm{Alexander~J.}\ \lnm{Wilson}}

\address[id={aff-swri}]{Southwest Research Institute, 1050 Walnut Street Suite 300, Boulder, CO 80302, U.S.A.}
\address[id={aff-swri2}]{Southwest Research Institute, 9503 W. Commerce, San Antonio, TX 78228, U.S.A.}

\runningauthor{G.T. Laurent et al.}
\runningtitle{WFI for PUNCH}

\begin{abstract}
We describe the design, hardware integration, and calibration performance of the Wide-Field
Imager (WFI) instruments for the Polarimeter to Unify the Corona and Heliosphere (PUNCH)
mission. The WFI instruments are a trio of visible-light heliospheric imagers that, 
together, view the outer corona and solar wind from under 3.5$^\circ$ to over 47$^\circ$
from the Sun, 
via sunlight that is Thomson-scattered from free electrons. In flight, the WFIs are arranged so
that their collective fields of view form an approximately symmetric trefoil on the sky,
comprising three circular-truncated square fields spaced 120° apart in position angle. 
The WFIs work with the NFI instrument, described elsewhere, to implement the full PUNCH 
field spanning all solar position angles, at elongations from 1.5° to 47° from disk 
center. WFI is implemented using dioptric (lens) optics and deep multi-stage baffles 
that attenuate solar, planetary, and lunar stray light sufficiently for ground processing 
to reveal the faint signal for the primary science. WFI measures both total brightness 
(tB) and polarized brightness (pB), via an on-board polarizing filter wheel (PFW) and 
charge-coupled device (CCD) camera that share a common design with those of the NFI 
instrument.
\end{abstract}
\end{frontmatter}

\section{Introduction}\label{S-Introduction}

\begin{figure}
    \centering
    \includegraphics[width=1.0\linewidth]{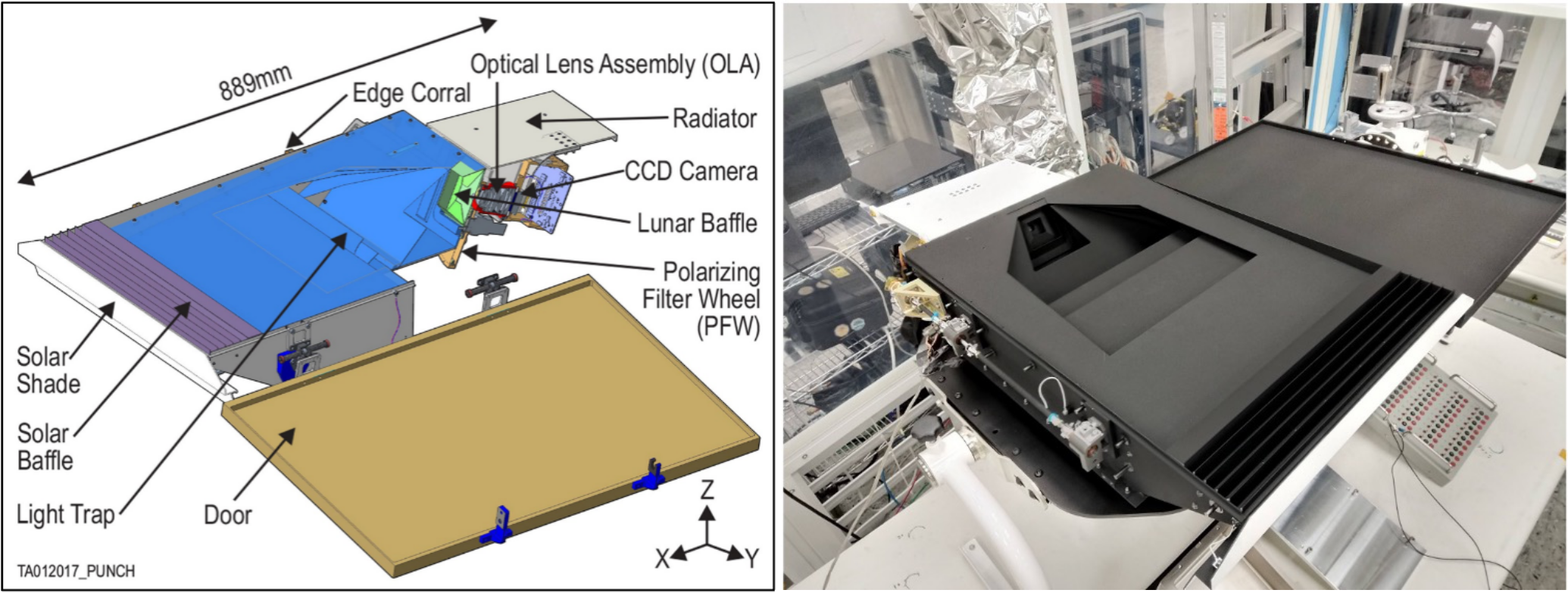}
    \caption{The PUNCH Wide Field Imager (WFI) instrument is a broadband visible-light imaging system. It
    produces images of the inner heliosphere using a three-stage baffle assembly with solar baffle, light trap, 
    and lunar baffle components, a polarizing filter wheel, optical lens assembly, and a 
    charge-coupled device detector operated in frame-transfer mode.  Left: schematic overview shows major 
    subsystems. Right: WFI-1 flight unit, photographed during integration, reveals stray-light minimizing features
    of the baffle assembly.}
    \label{fig:wfi}
\end{figure}

The Wide-Field Imagers (WFIs) are the three identical heliospheric imagers that 
are part of the Polarimeter to Unify the Corona and Heliosphere mission 
\citep[PUNCH;][]{DeForest_etal2026}. WFI (Figure 
\ref{fig:wfi}) is a deeply-baffled, polarizing, wide-field, broadband visible 
light imager.  The WFIs work together with the PUNCH ground data reduction
pipeline to produce regular-cadence images of the solar wind via the faint 
traces of sunlight Thomson-scattered by free electrons. 

\begin{figure}
    \centering
    \includegraphics[width=0.5\linewidth]{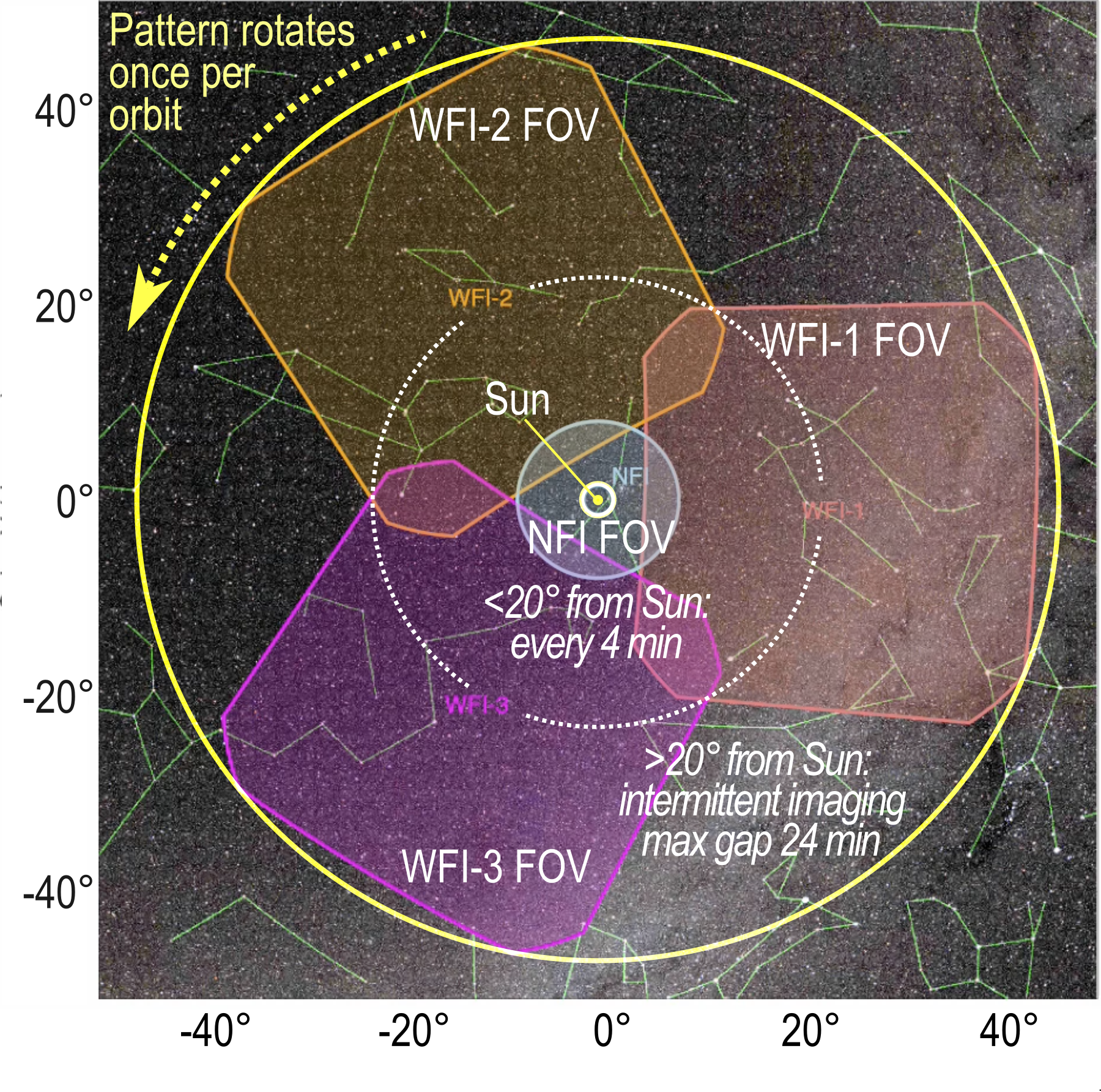}
    \caption{Each WFI instrument has an approximately square field of view subtending  44$^\circ$ of sky; the three WFIs together sweep out the full over-90$^\circ$-wide outer portion of the PUNCH mission field of view. (\textit{Image: DeForest et al. 2026, Fig. 2)}}
    \label{fig:fov}
\end{figure}

The PUNCH mission images plasma and space-weather phenomena throughout
the Sun's outer corona and the inner heliosphere. Four meter-class 
satellites (“microsats”) 
each carry one primary instrument and work together to produce images
across an extremely wide field of view spanning over 90°, centered 
on the Sun.  The three 
Wide-Field Imagers together capture the 
outer portion of the full PUNCH FOV; a separate Narrow-Field Imager \citep[NFI;][]{colaninno_2025}
captures the inner portion of that FOV. The WFI and NFI instruments are
operated to maintain overlap between their individual FOVs, and are operated 
synchronously, so that their data can be combined on the ground. Figure \ref{fig:fov} 
shows the WFI and NFI FOVs for a single observation. The 
science objectives, mission design, observing sequence, and data products are 
detailed by \citet{DeForest_etal2026}.  In this article, we describe technical 
aspects of the WFI instrument in particular.

Heliospheric imaging was pioneered by \citet{jackson_leinert_1985}, who  
constructed and isolated highly sensitive, if low resolution, image data products
from three-photometer timeseries 
data acquired by the spinning Helios spacecraft \citep{pitz_etal_1976}. The first 
dedicated heliospheric imager, the Solar Mass Ejection Imager 
\citep[SMEI;][]{eyles_etal_2003}, 
used three deeply baffled cameras with quasi-linear fields of view, to sweep 
out the entire $4\pi$ steradians of sky once per orbit of its host spacecraft
(Coriolis, 
in Sun-synchronous Earth orbit at roughly 800~km altitude). The quasi-linear images 
were recombined on the ground, yielding all-sky maps that were used for CME tracking 
and solar wind tomography \citep[e.g.,][]{tappin_etal_2004,jackson_etal_2006} among 
other applications. 
Shortly thereafter, 
the Solar-Terrestrial Relations Observatory \citep[STEREO;][]{kaiser_etal_2008} carried the 
Sun-Earth Connection Coronal and Heliospheric Imaging suite 
\citep[SECCHI;][]{howard_etal_2008}, which included two pairs of 
heliospheric imagers \citep[HI1 and HI2;][]{eyles_etal_2009}, 
one pair on each of two spacecraft. The HI1 and HI2 cameras shared a single 
directed-corral style baffle with a multi-edge ``horizon'' and deeply-vaned 
light trap; each imaged a wide swath of sky with a full 2-D image plane and 
dioptric focusing 
system.  The pair covered roughly 45$^\circ$ in solar position angle out to a 
maximum 
elongation above 90$^\circ$. Several newer instruments 
build on the success of the STEREO/SECCHI-HIs; these 
include the Solar Orbiter Heliospheric Imager \citep[SOLOHI;][]{howard_etal_2020},
the Wide-field Imager for Solar Probe \citep[WISPR;][]{Vourlidas_etal2016}, 
and the in-development Vigil/HI instrument \citep{Tappin_etal_2023}. 

The overall WFI design echoes those of STEREO/HI, SOLOHI, and WISPR: light first 
passes over a baffle system at the front of the instrument, which dominates the 
overall form factor. Light then passes through an external entrance pupil, a motorized polarized 
filter wheel (PFW), and a custom dioptric optical lens assembly (OLA), before finally 
coming to a focus on a CCD image sensor. Figure \ref{fig:wfi} highlights each of these 
major components and subsystems. The design architecture of WFI parallels that of 
NFI to simplify subsystem-level design and ensure common operations across PUNCH. 
The design incorporates several common elements that are identical between the 
instruments, including the CCD camera and associated readout electronics, PFW, 
instrument door actuators, and thermal components (heaters and thermistors). The 
instruments and spacecraft are interchangeable, with the only control differences 
consisting of flight software tables that specify the instrument control signals and 
sensor ranges.

WFI rests on three titanium flexure feet arranged as a kinematic mount and designed 
to thermally isolate WFI from the spacecraft. A one-shot spring-loaded door, powered 
by a high-output paraffin pin puller, maintains baffle cleanliness during launch. The 
WFI camera is cooled using a passive radiator that is mounted below the baffle plane 
to minimize scattered light. Active trim heaters throughout maintain operational 
temperatures. Additional survival heaters ensure that the subsystems stay above their 
survival limits if the instrument is unpowered. 

WFI differs from other modern heliospheric imagers in two important respects:  (1) 
WFI includes a two-bounce ``lunar baffle'' intended to reduce the effect of lunar 
glint, since WFI operates inside the Moon’s orbit; and (2) WFI measures polarization 
of light from the corona and solar wind. Polarization is used by PUNCH to infer the 
3-D structure of the corona and of bright features in the solar wind. An overview of 
the theory is in DeForest et al. (2026) and more detailed treatments are 
in \citet{Gibson_etal_2026}, \citet{deforest_etal_2013}, and \citet{tappin_howard_2009}.

Each WFI has a useful FOV that is 44$^\circ$ square, truncated 
by a 55$^\circ$ circle, on the celestial sphere, providing margin against a nominal 40° square 
truncated by a 50$^\circ$ circle. The optical focal length is 35.2mm, operated at f/3.2 (11 
mm aperture), yielding a plate scale of approximately 1.47$'$ per 15 $\mu$m pixel 
(varying slightly across the field). The 
observing passband is 450~nm to 750~nm (broadband visible light). The dioptric 
imaging system is 
based on modern eyepiece designs, operated in reverse to produce a real image with 
the collimated light arriving from celestial objects.

\begin{figure}
    \centering
    \includegraphics[width=0.75\linewidth]{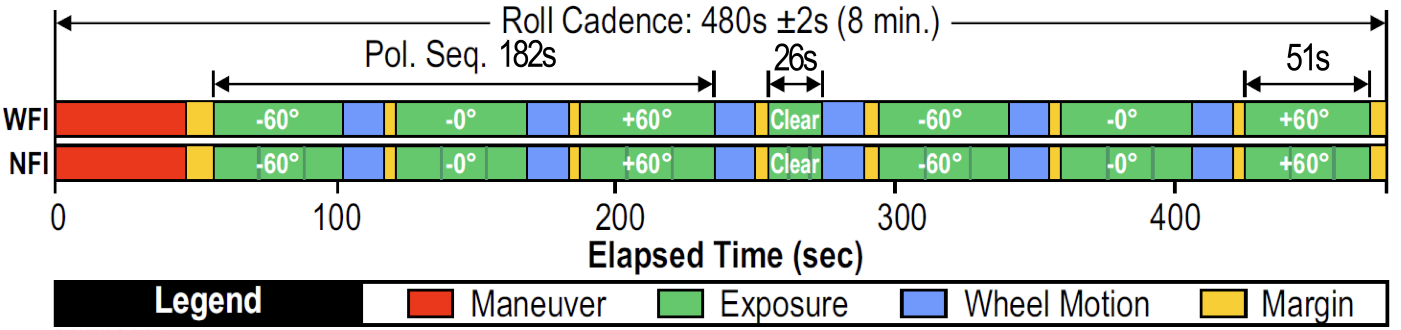}
    \caption{The PUNCH observing sequence supports seven exposures (including two complete M,Z,P 
    polarization sequences)
    during each 8-minute orbital observing window. The sequences are synchronized across all four 
    primary instruments. (\textit{Image: DeForest et al. 2026, Fig. 8)}}
    \label{fig:observing-sequence}
\end{figure}

The PUNCH observing sequence consists of near-identical sequences for WFI and NFI 
during each 8-minute orbital observing period. The instruments acquire two 
polarization sequences at three angles each and one unpolarized image taken with a 
clear filter. The polarized images are single 51-second exposures and the unpolarized 
(clear) images are single 26-second exposures. The polarization sequences use the 
(M,Z,P) formalism described by DeForest et al. (2022), and fully characterize the 
linear polarization state of incoming light using polarized filters at angles of -60$^\circ$,
0$^\circ$, and +60$^\circ$ relative to instrument vertical. Figure \ref{fig:observing-sequence} shows the 
coordinated observing 
program for the PUNCH mission, which makes polarized image sets every four minutes. 
The WFIs are designed and calibrated to be as interchangeable as possible: 
their intended use is to create the 90$^\circ$ diameter PUNCH mosaic trefoil data products.  

\begin{figure}
    \centering
    \includegraphics[width=0.67\linewidth]{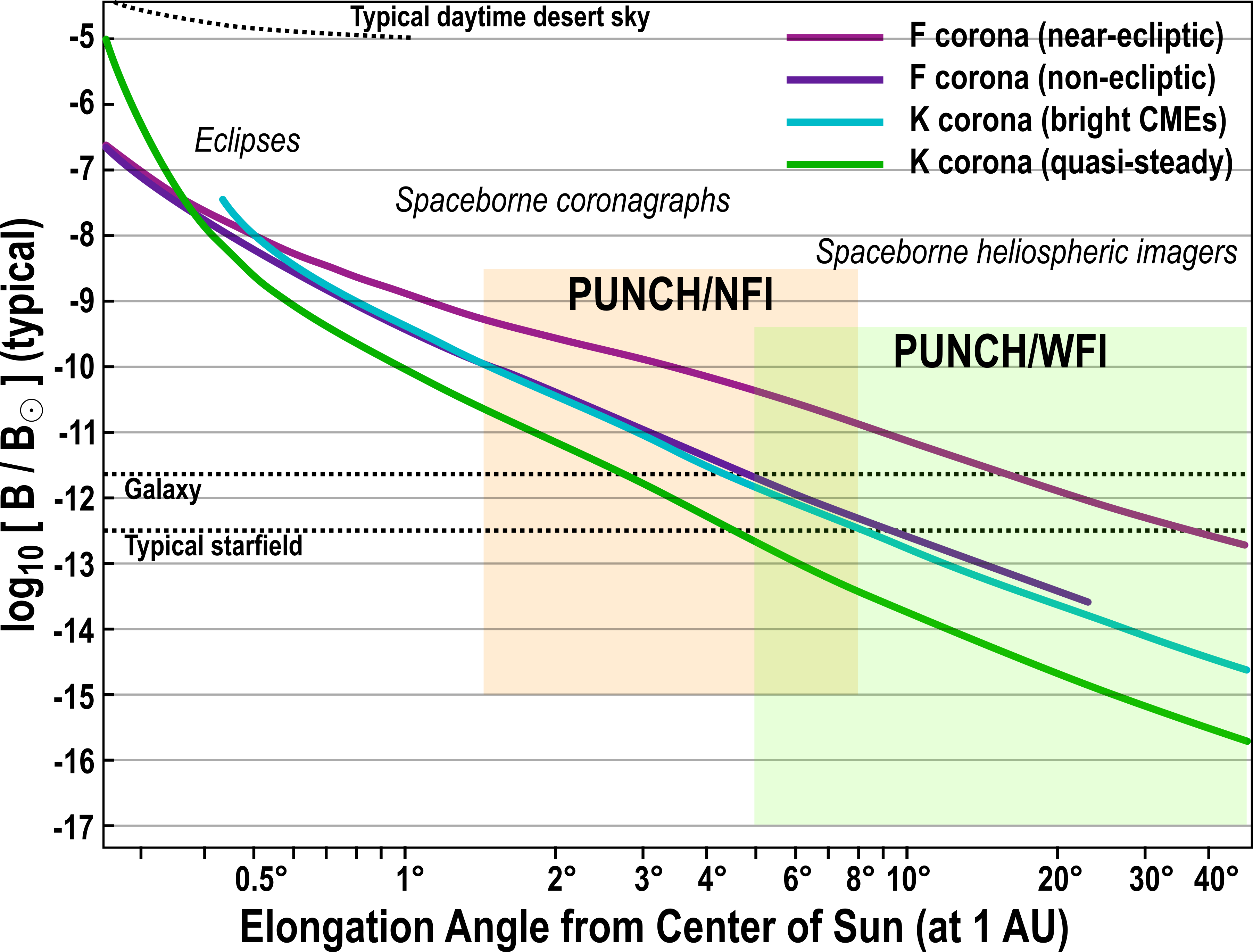}
    \caption{The challenge of heliospheric imaging is summarized in this typical-brightness plot, showing
    radiances of various celestial sources in units of $B_\odot$, the disk-averaged solar photospheric 
    radiance.  At elongation angles above 20$^\circ$ from the Sun, the faint dynamic signal from the solar 
    wind is roughly 1,000 times fainter than the near-ecliptic F corona or the background starfield.
    (\textit{Image: DeForest et al. (2026), Fig. 10})}
    \label{fig:brightnesses}
\end{figure}

Heliospheric imager instruments in general, and WFI in particular, should be considered
as part of an extended imaging system that includes significant post-processing elements
for calibration and background separation. Published PUNCH/WFI images are heavily processed by
the mission data-reduction pipeline both for calibration and also for removal of multiple
backgrounds that are, themselves, much brighter than the signal under study.  Calibration
steps include corrections for photometry, point-spread function
including both static and dynamic (spacecraft pointing) effects, optical distortion, 
and image co-registration among others.  Background removal steps include 
scrubbing for cosmic rays, detector artifacts, and streaks caused by other satellites;
removal of terrestrial stray light, solar stray light, geocorona, zodiacal light or ``F corona'';
minimization of glint from bright celestial objects including the Moon; and characterization and
removal of the starfield itself.  The challenge
of heliospheric post-processing has been detailed by many authors including, e.g., 
\citet{jackson_leinert_1985},
\citet{eyles_etal_2003}, \citet{jackson_etal_2011},  \citet{deforest_etal_2011}, 
\citet{stenborg_etal_2018}, and \citet{howard_etal_2020}, and is summarized by the 
radiance-vs-elongation plot in Figure \ref{fig:brightnesses}, which shows the factor-of-1,000
difference between the brightness of the zodiacal light and the solar-wind
signal, more than 30$^\circ$ from the Sun itself.  
The PUNCH data-reduction pipeline
is described in a separate article in this Special Collection 
\citep{hughes_etal_2025},
and is continuously published as open-source software together with 
inline documentation \citep{punchbowl}.

\begin{figure}
    \centering
    \includegraphics[width=1.0\linewidth]{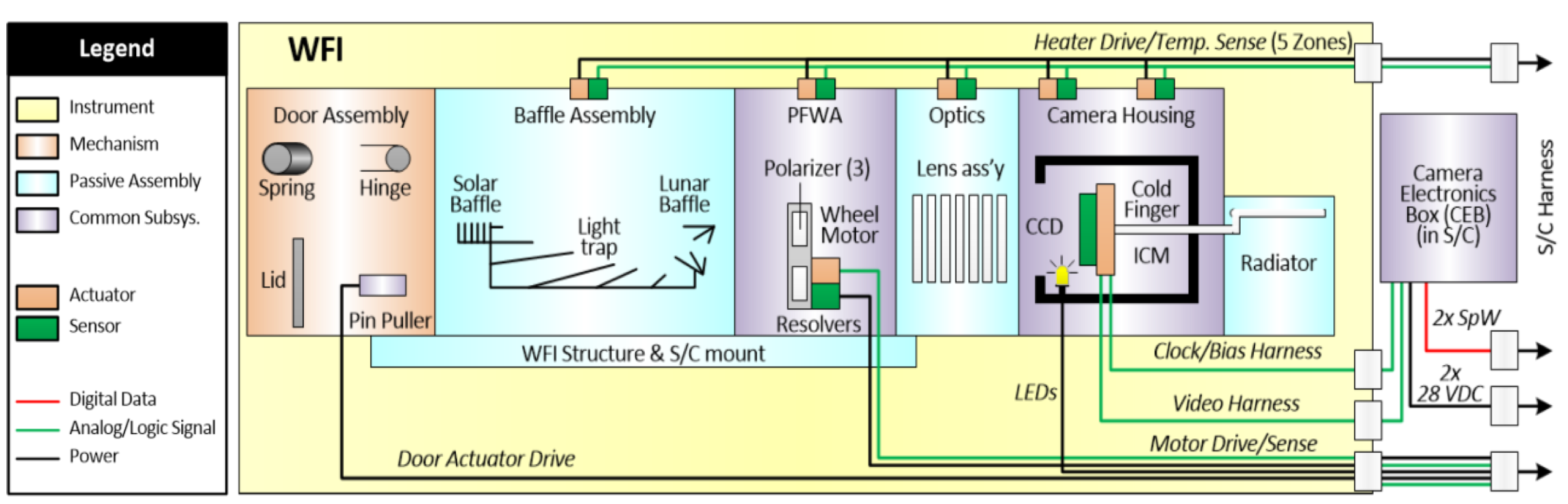}
    \caption{Functional block diagram of the WFI instrument, with common subsystems identified, shows simple overall structure and parallelism to PUNCH/NFI \citep{colaninno_2025}.}
    \label{fig:block-diagram}
\end{figure}

The overall WFI design is conceptually straightforward and is captured in the instrument block diagram
(Figure \ref{fig:block-diagram}).  The design is intentionally highly parallel to that of NFI 
\citep{colaninno_2025}: the instruments differ in physical form and in particular observing sequence, 
but have identical electrical interfaces and several common subsystems.
In the following sections, we describe: (Section \ref{S-optical}) optical 
design and performance; (Section \ref{S-mechanical}) mechanical design and performance; 
and (Section 
\ref{S-thermal}) thermal design and performance. A short discussion and conclusion
(Section \ref{S-discussion}) covers WFI’s as-flown role within PUNCH and relative to 
other heliospheric imagers, and anticipated 
relevance to solar wind science.  

\section{Optical Design and Performance}\label{S-optical}

The purpose of WFI is to carry out deep-field imaging of a large field of view (FOV); this drives the
optical design.  The optics and overall instrument design are dominated by  
stray light rejection requirements and the large FOV, which impose  
a particular baffle design (Section \ref{SS-baffle}). The large FOV drives placing the polarization 
resolver, a polarizing filter wheel (PFW), in front of the primary optics for 
uniformity (Section \ref{SS-PFW}). 
Imaging is accomplished by dioptric optics in an Optical Lens Assembly 
(OLA; Section \ref{SS-OLA}). The image detector is a CCD detector 
operated in frame-transfer mode, with no moving parts (Section \ref{SS-Camera}). Data are
processed on-board to reduce total data volume while preserving photometry (Section \ref{SS-data-processing}).
The instrument was tested end-to-end during integration (Section \ref{S-testing}).

\subsection{Baffle}\label{SS-baffle}

\begin{figure}
    \centering
    \includegraphics[width=1.0\linewidth]{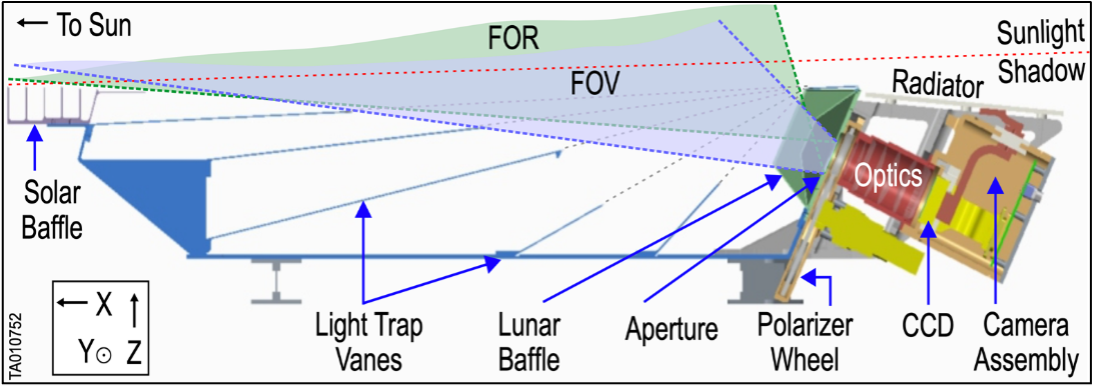}
    \caption{Cross-sectional view of the WFI baffle shows its three-part design: solar baffle, light trap, and lunar baffle. The baffle is surrounded by a sidewall that, in conjunction with the host spacecraft roll program, eliminates direct scattered light from the nearby Earth.}
    \label{fig:fov-for}
\end{figure}

The WFI baffle (Figure \ref{fig:fov-for}) has three major parts: a ``solar baffle'' to reduce Fresnel 
diffraction into the instrument; a ``light trap'' to reduce the effect of glint and other stray light 
sources; and a ``lunar baffle'' to reduce the angular field of regard (FOR) and suppress moonlight.
The fundamental design is a planar corral with a leading-edge 
multiple-vane baffle (the ``solar baffle''). The multiple vanes overcome 
Fresnel diffraction from incident sunlight. In the ray approximation, the entire instrument 
remains within the umbral shadow cast by 
the first vane of the solar baffle. The instrument aperture is nestled inside a deeply 
vaned ``light trap'' that prevents glint and absorbs moonlight incident from the +Z half-space (with the coordinate system shown in Figure 6). 
The aperture itself is protected by a conventional two-bounce ``lunar baffle'' that reduces 
the field of regard (FOR)\footnote{Readers are reminded that an instrument's ``field of regard'' (FOR) is the range of angles from which incident rays can impinge directly on the optics. This is generally wider than the field of view (FOV), which is the range of angles from which incident rays are focused by the optics into a usable image.}. 
The FOV and FOR are restricted in the lateral and upward directions by the lunar baffle, 
and in the 
forward direction by the solar baffle. The light trap and lunar baffle are both designed such 
that only the dark (interior) side of each baffle vane is visible to the aperture. The light trap and 
lunar baffle vanes are sharpened to reduce glint from their edges. The space between each pair of vanes 
forms a deep cavity: specular or near-specular scatters of incident light cause the light to be scattered 
more deeply into the cavity, attenuating the beam’s power by multiple interactions 
with the blackened surface 
of each vane.

\subsubsection{Solar baffle}\label{SSS-solar-baffle}

The WFI baffle vane is an extrusion along the spacecraft Y axis, of a 3-D design in the spacecraft XZ 
plane. The baffle is designed for an 11~mm circular or square aperture, with a clear (unvignetted) field 
of regard beginning 5.5$^\circ$ from Sun center, and a minimum vignetting function of 0.2 
(80\% vignetted) 
at 5.0$^\circ$ from Sun center. The overall length in the X direction is 620~mm from the leading edge to 
the center of the aperture, balancing a design trade between long baffle throw and compact spacecraft size.

\begin{figure}
    \centering
    \includegraphics[width=1.0\linewidth]{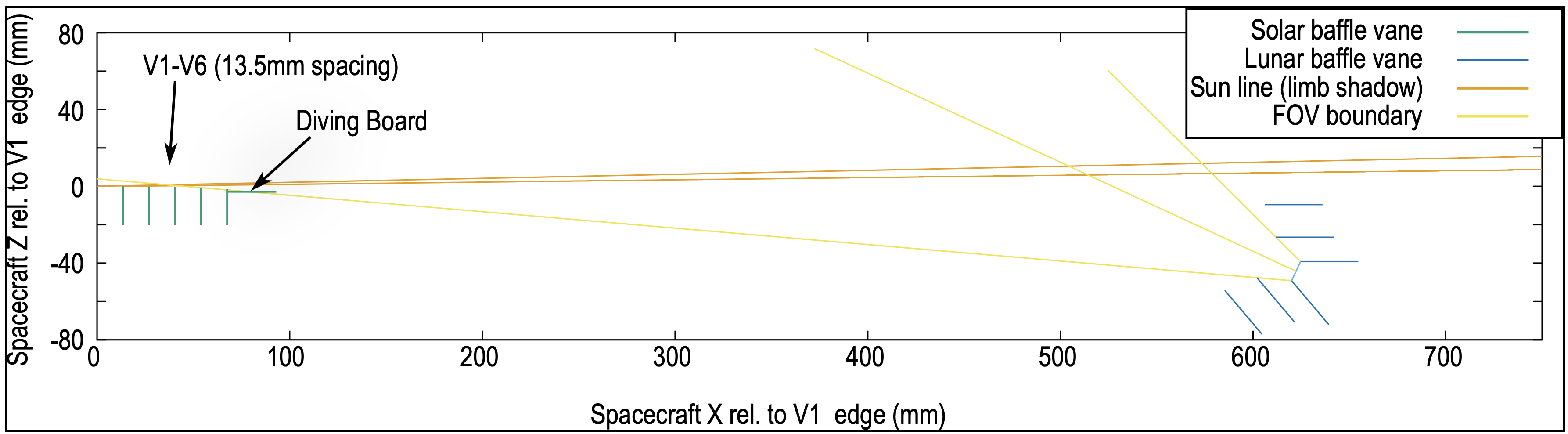}
    \caption{Parametric 2-D design produced by the WFI design software places the solar and lunar baffle edges in the notional instrument XZ plane for extrusion along Y and ultimate fabrication.  The instrument overall length, solar location, FOV boundary angle, leading-edge vane spacing, vane count, and diving board length are parameters of the design software, which outputs a sketch design like this one and also a predicted scattering performance.  The shadow line rises at a slope of 0.6$^\circ$ from the origin. The aperture is at right, at the center of the lunar baffle.}
    \label{fig:vane-design}
\end{figure}

The WFI solar baffle comprises 6 vanes (Figure \ref{fig:vane-design}), 
numbered V1-V6. Following the STEREO/HI design 
\citep{Halain2007,Halain2011}, 
the vanes are chamfered on the ``dark'' (anti-Sun, spacecraft -X) 
side to present 
a more sharpened edge and a flat vertical surface to incoming sunlight. The vane depth is set to 20~mm, 
chosen to make the trench aspect ratio close to (and greater than) unity, but is not a driver of the 
design. The overall envelope of the edges is approximately cylindrical, with a fixed amount of bend
at each vane.  The edges (tops in 2-D) of adjacent vanes form planes (lines in 2-D) 
with a dihedral angle $\Delta\Theta$ of 0.6$^\circ$ between adjacent planes formed by any three
consecutive vanes. The vanes are separated horizontally by a distance $\Delta L$ of 13.5mm. The 
first two vanes have equal 
height relative to the instrument, and subsequent vanes are slightly shorter to accommodate the 
increasing absolute angle. The six vanes form five individual planes, and the last plane (formed by the 
V5 and V6 edges) is canted 2.4° from horizontal. The baffle is 
designed to be operated with Sun center 0.85$^\circ$ below the -X direction, so that the 
limb of the Sun is 
approximately 0.6$^\circ$ below the -X direction (but see the Discussion in Section 5).  

The last vertical solar vane V6 is followed by a “diving board” V7 that is 
cantilevered out from the solar baffle and forms a hybrid solar baffle / light trap vane. 
The diving board is 
sized and placed to slightly vignette the aperture at the innermost edge of the FOV, and is therefore
part of the solar baffle. The purpose of the diving board vane is to prevent moonlight from 
scattering directly off the rear face
of the solar baffle into the aperture during the gibbous lunar phases; it also prevents any glint 
from the rear of the spacecraft from doing the same. 

Vanes V1-V7 have tight relative alignment tolerances and are therefore machined from a single block of
aluminum.  They are shimmed into place relative to the rest of the instrument during overall 
integration.

The exact placement of the solar baffle vanes is driven by Fresnel scattering of sunlight 
around the vanes; we therefore designed the vane count, placement, and spacing iteratively, 
with custom parametric software implementing the Sequential Plane Wave (SPW) approximation detailed by \citet{deforest_etal2025}. 
In the context of the WFI solar baffle, the SPW approximation yields the approximate formula:
\begin{equation}
    I_{obs} \approx I_0\left[\mathcal{M}\left(\Delta\Theta\sqrt{\pi\Delta L / \lambda}\right)\right]^n \times \mathcal{M}\left(\Delta\Theta\sqrt{\pi d/\lambda}\right) \times \mathcal{M}\left(\Delta\Theta\sqrt{\pi D/\lambda}\right)\,,\label{eq:SPW}
\end{equation}
where $I_{obs}$ is the observed intensity in shadow, $I_{o}$ is the incident light intensity, 
$\mathcal{M}()$ is the Fresnel integral described by \citet{deforest_etal2025}, $\Delta\Theta$ is the 
bend angle around each vane, $\Delta L$ is the distance between vanes, $\lambda$ is the wavelength of 
light, $n$ is the number of solar vanes, $d$ is the distance from the final solar vane to the ``diving 
board'' vane at the rear of the solar baffle, and $D$ is the distance from the diving board vane to the 
optics.

The design software yielded 
predicted stray light based on Equation \ref{eq:SPW}, numerically integrated over the 450-750 nm 
acceptance passband of the instrument in 1 nm increments with uniform weighting (‘white spectrum’ 
approximation). Rather than integrate over the angular size of the sun, we used the point source 
approximation, placing the point source at the worst location on the solar limb rather than at Sun 
center. Likewise, we considered the aperture to be the fully located at the worst location (highest point 
on the aperture line, in 2-D). These two approximations together yielded conservative predictions 
compared to full-aperture and full-disk integration, but for an ideal 
baffle geometry without mechanical tolerances applied. The software was rapid enough
to allow exploration of the design space, converging on the final design.

\begin{figure}
    \centering
    \includegraphics[width=0.5\linewidth]{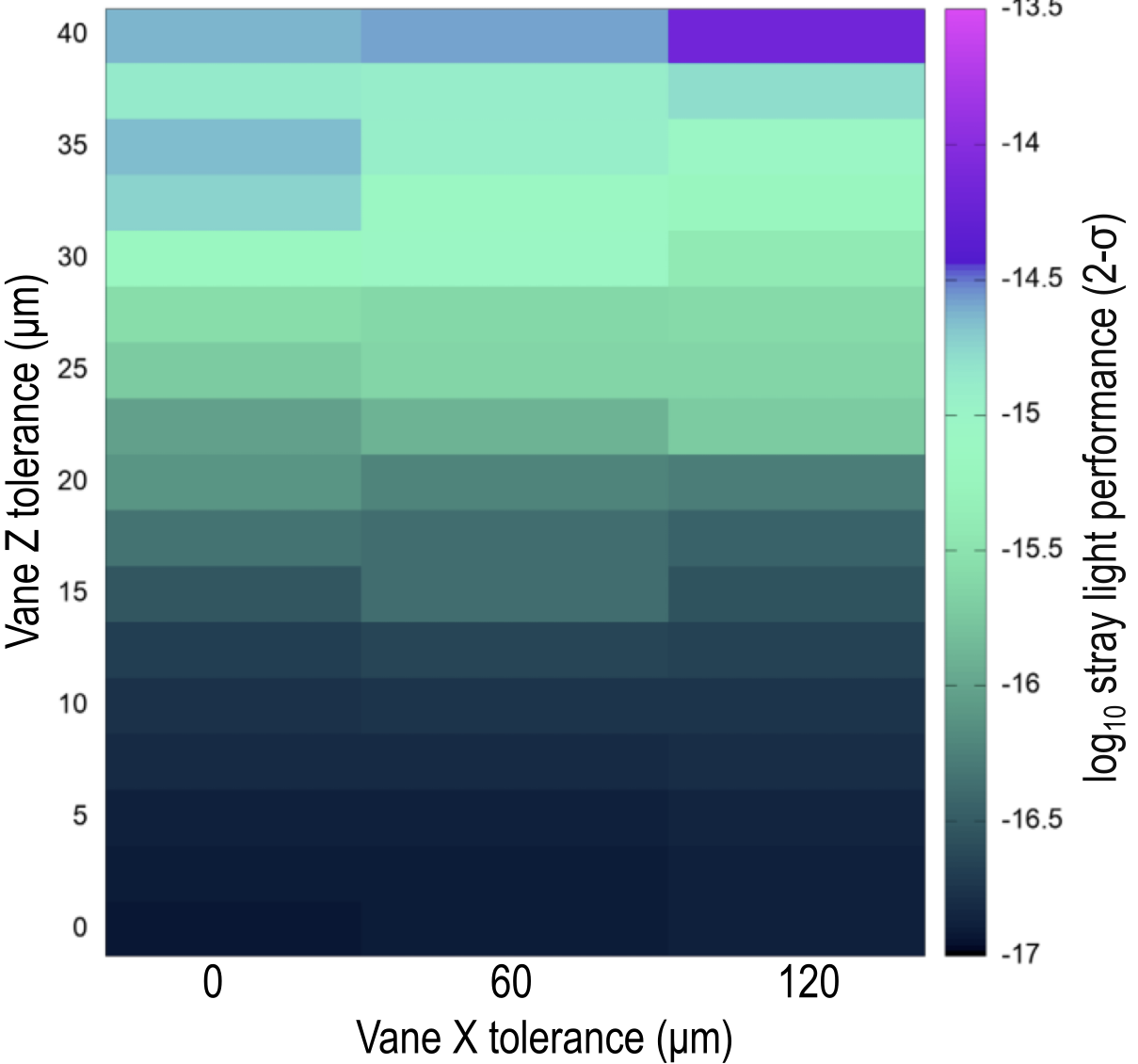}
    \caption{Monte Carlo simulation results yield forecast stray light performance vs. solar vane edge positional tolerance and show that stray light is robust against vane tolerancing up to 30~$\mu m$ in the Z direction (height) and 120~$\mu m$ in the X direction (spacing).}
    \label{fig:tolerancing}
\end{figure}

We used a Monte Carlo approximation approach to develop mechanical tolerances for 
fabrication.  In this step, we used the SPW formula and randomized perturbation of the calculated
ideal vane placements to determine expected-value performance of the baffle under different mechanical 
tolerances. For this step, we considered the aperture as representative top, middle, and bottom regions 
rather than a single point.  X and Z tolerances were treated separately: for a given tolerance, every vane 
edge was perturbed by one sample of a Gaussian random variable centered on 0, with variance equal to the 
tolerance.  The diving board tolerances were doubled compared to the nominal tolerance of the solar 
vanes, accounting for the mechanical disadvantage of the cantilever.  We calculated rejection tolerances 
for an ensemble of 400 baffles, calculated the mean $\mu$ and variance $\sigma$ of the ensemble, and 
reported the prediction scattering performance as $\mu + 2\sigma$. Figure \ref{fig:tolerancing} shows the 
predicted scattering performance vs. independent X and Z tolerances on vane placement.

Individual as-built WFI solar baffles were measured and as-built performance was forecast using the 
same code.  All were well within specification.  Actual Fresnel scattering was measured for one flight
WFI solar baffle; that test is described in Section \ref{SS-testing}.

\subsubsection{Light trap}\label{SSS-light-trap}

The light trap is designed to prevent glint or multiple internal corral reflections from entering the 
aperture. It comprises six sharpened knife-edge vanes, V7 (the diving board, which is mechanically part 
of the solar baffle) through V12. Its vanes are sharpened and are oriented so that only the sharpened 
edge and dark underside of each vane is visible to the aperture. The planes of V7 through V12, extended 
beyond the edges of the vanes themselves, meet just above the top vane of the Lunar Baffle (Figure 
\ref{fig:fov-for}).

The vanes V7-V10 are inside the FOR of the lunar baffle and can scatter light directly into the baffle. 
V11 and V12 are outside the FOR of the lunar baffle and edge scatter is negligible compared to direct 
moonlight effects. The attenuation of moonlight is proportional to the reflectivity of the edge, the 
surface area of the edge, and the apparent size of the aperture from the edge itself compared to the 
hemisphere into which moonlight is scattered. The apparent size of the aperture varies from edge to edge 
but the average distance is 350~mm, for an average subtended solid angle of roughly 1~mSR relative to
each potential scattering source on the edge. The sharp 
edges are treated with the same black surface treatment as the rest of the vane, minimizing direct glint 
and further attenuating moonlight scatter.

The surface area that can receive moonlight and shine directly into the aperture from each vane is 
approximately one edge radius times the length of the vane, so $A_{edge}\approx (5~\mu m)
(200~mm)=1.1\times 10^{-5}~m^2$, illuminated by moonlight.  That edge subtends a solid angle
$\Omega_{edge}$ of approximately $1\times 10^{-4}$~SR relative to the aperture.  Neglecting absorption 
from the surface treatment, lunar intensity is conserved by scattering off the vane and therefore the 
radiance of the vane is reduced by a factor $\alpha_{sc}=\Omega_{moon}/2\pi=10^{-5}$. The surface 
treatment reflectivity budgeted value was 0.05 for this calculation, for an overall edge scatter of:
\begin{equation}
    I_{edges} = I_{\odot} n_{edges}\frac{\Omega_{moon}}{2\pi}\frac{\Omega_{edge}}{2\pi}R_{surf}\frac{I_{moon}}{I_{\odot}}=3\times10^{-16}I_\odot\,,\label{eq:light-trap-scatter}
\end{equation}
which is comparable to the computed Fresnel scattering.

Indirect scatter from vane surfaces into the aperture is computed to be 1-2 orders of magnitude fainter than from the vane edges.

\subsubsection{Lunar baffle}\label{SSS-lunar-baffle}

The lunar baffle is a conventional two-bounce baffle, with three apertures: the primary aperture of the 
optics, an intermediate aperture (L2 vanes), and an outer aperture (L1 vanes). The geometry is 
constructed such that the L1 vanes are obscured from the aperture by the L2 vanes in the ray 
approximation. Because of the geometry, vane surface scattering is negligible (as with the light trap) 
and dual edge scatter 
dominates overall performance.

Taking the same approximation as for the light trap, the exposed edge area of L1 that can 
scatter onto L2 is roughly $(30~mm)(5~\mu m)$, or $150\times 10^{-9}~m^2$. Because of the 
sharper angle of attack compared to the light trap vanes, the effective scattering area is 
reduced by another factor of 3, to $50\times10^{-9}~m^2$. Light scattered from this area 
is lost unless it impinges on L2. L2 has half the area of L1 and averages a distance of 
30~mm from L1, and therefore, averaged over the edge of L1, subtends a solid angle of 
$50~\mu SR$ for an overall attenuation from L1$\rightarrow$L2 of $2\times10^{-5}$. A second 
Lambertian scatter into the aperture, which subtends 1 steradian from L2, yields a further 
attenuation of approximately $\pi$, for an overall effective area (for scatter into the 
optics) of $3\times 10^{-13} m^2$. The geometric area of the aperture is $1\times 
10^{-4}~m^2$, for an overall attenuation of $3\times 10^{-9}$, neglecting surface 
treatment, which imposes another factor below $10^{-2}$. Thus moonlight is attenuated by an overall
factor of order $10^{-11}$ by the lunar baffle, yielding a stray light intensity less than
$10^{-16} I_{\odot}$, which is below the computed Fresnel scattering and light trap stray light.

\subsection{Polarizing filter wheel}\label{SS-PFW}

Common to all four primary PUNCH instruments (each WFI and NFI) is a 5-position polarizing filter wheel (PFW) assembly. 
The PFW is described in detail by \citet{colaninno_2025}. While the PFWs are mechanically identical for both WFI and 
NFI, the filters are oriented differently to produce the same on-sky polarization angles despite being mounted in 
different directions in the instrument optical trains.  For the WFI implementation, the actuator is mounted 
to the mechanical structure between an external pupil (Section 2.3) and OLA entrance aperture (rotating clockwise with 
respect to the spacecraft +X direction). The filter wheel holds three polarizing filters mounted at 60° relative angles 
(-60$^\circ$, 0$^\circ$, +60$^\circ$) to capture the full polarization state as in \citet{deforest_etal2022b}, plus an 
opaque dark position and a clear (unpolarized) filter. The polarizers are nanowire type, chosen for their very broad 
wavelength range, insensitivity to incidence angle, and radiation tolerance.

\subsection{Optical Lens Assembly}\label{SS-OLA}

\begin{figure}
    \centering
    \includegraphics[width=0.75\linewidth]{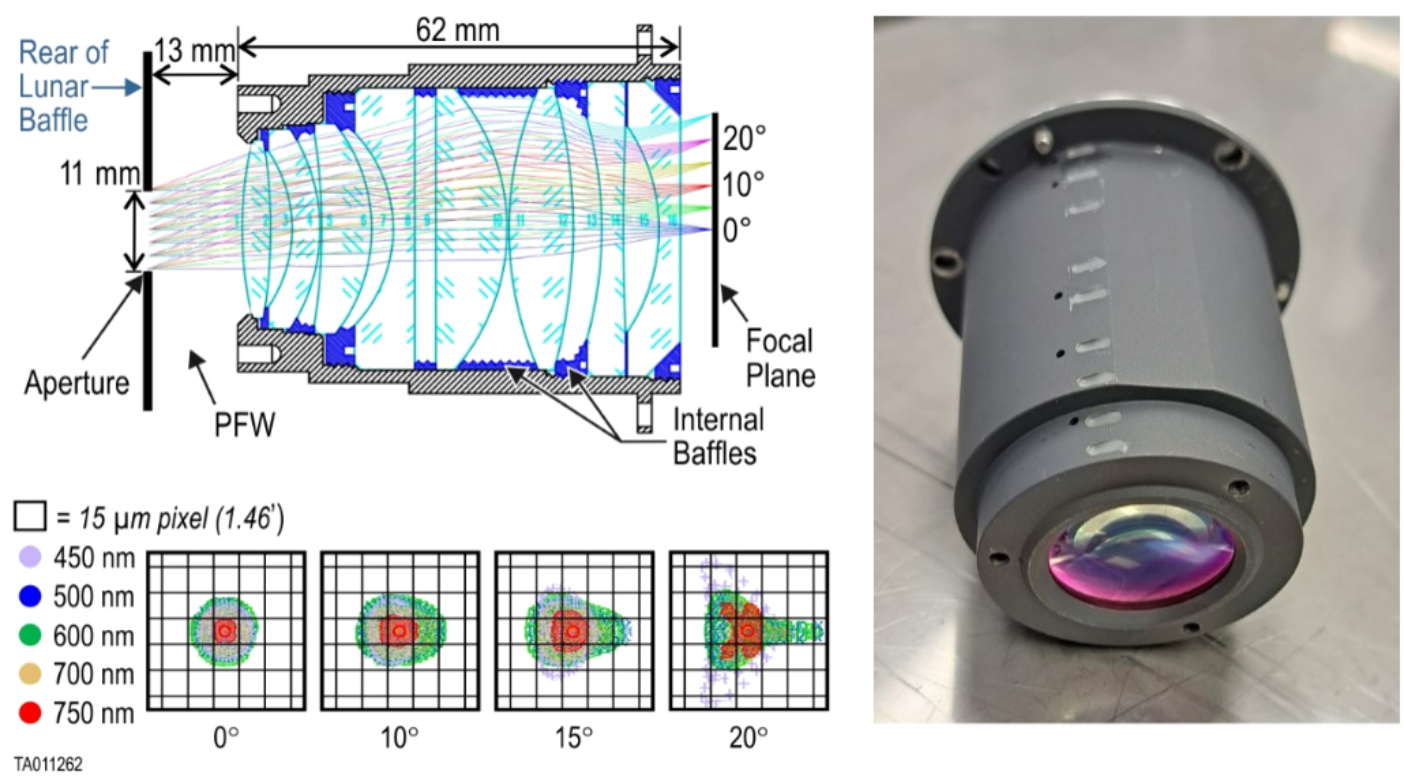}
    \caption{The WFI Optical Lens Assembly (OLA) is an 8-element design based on the Nagler Type 2 eyepiece, operated in reverse to form a real image of the celestial sphere. The optical ray trace shows good uniformity and achromaticity across the field. The optics are mounted in a baffled titanium tube, and potted into place with sealant.}
    \label{fig:OLA}
\end{figure}

\begin{figure}
    \centering
    \includegraphics[width=0.67\linewidth]{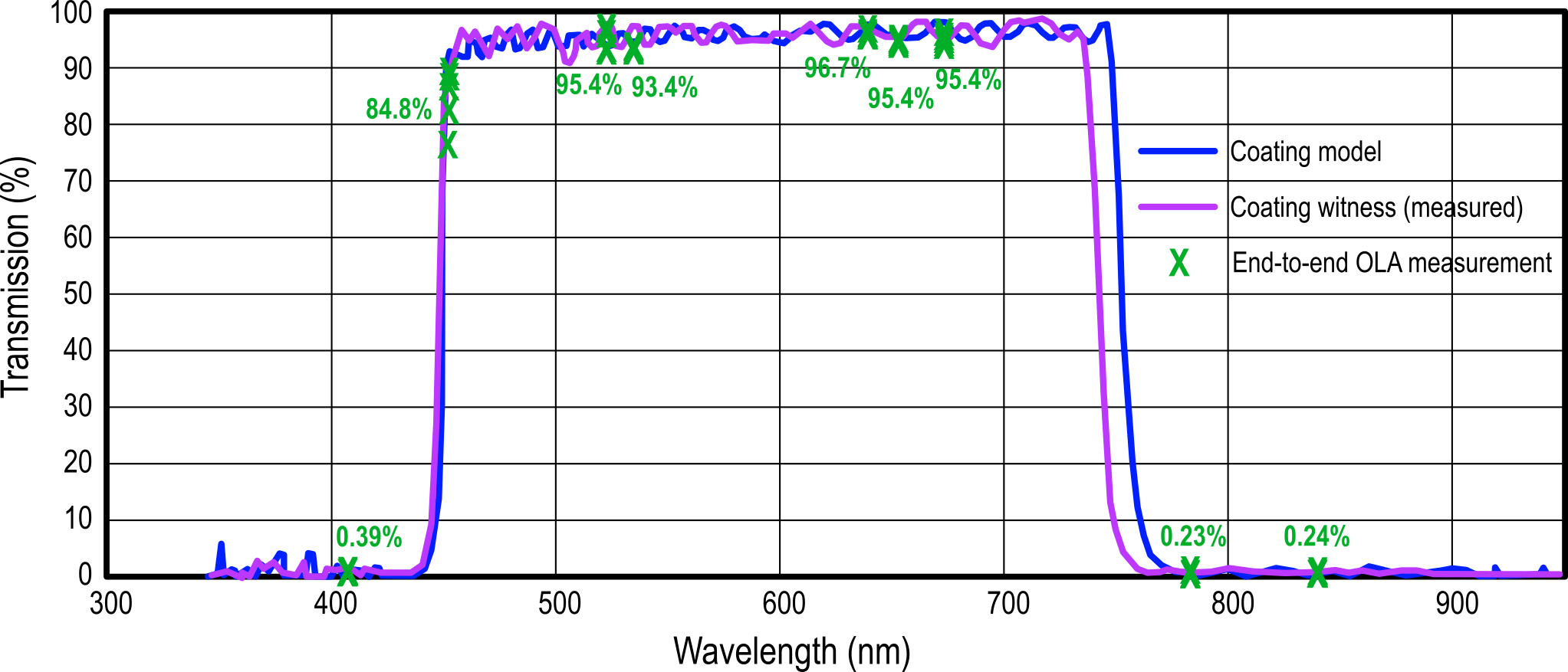}
    \caption{Transmission profile (modeled vs. measured) of the OLA bandpass coating agrees with end-to-end transmission profiles of the complete OLA. Five separate flight model OLAs were tested at at nine wavelengths; ensemble-average transmission is given at each wavelength.}
    \label{fig:passband}
\end{figure}

The Optical Lens Assembly (OLA) is a variant of the Nagler Type 2, 8-element eyepiece design
\citep{Nagler1988}, created by Al Nagler himself and fabricated by Tele~Vue Optics.  
It comprises 8 optical elements mounted in a 
stepped titanium lens tube (Figure \ref{fig:OLA}), potted with flexible, vacuum-compatible sealant.  
The glass materials include conventional flint, crown, and high dispersion rare-earth glasses.
Titanium was selected for the enclosure for its good thermal expansion coefficient 
match with most  optical glasses, leading to a very wide survival temperature range of $\pm 60$~C.  
The eyepiece optical design is used in reverse configuration, accepting collimated light and focusing it
onto a detector plane behind the lens system.  The design has an 
external pupil, which is placed at the aperture of the WFI lunar baffle to reduce stray light.  
The ``eye relief'' 
between the pupil and first lens is used to house the PFW 
(Section \ref{SS-PFW}).
The optics are modified from commercial Nagler eyepiece designs so that the real image is 
outside the physical limits of the 
lens train, allowing capture by a physical detector without re-imaging. The passband of 450-750~nm is defined by a
bandpass interference filter on the leading edge. The overall transmission coefficient is 96\% in-band and averages 
less than 0.3\% out of band (Figure \ref{fig:passband}).

\begin{figure}
    \centering
    \includegraphics[width=0.5\linewidth]{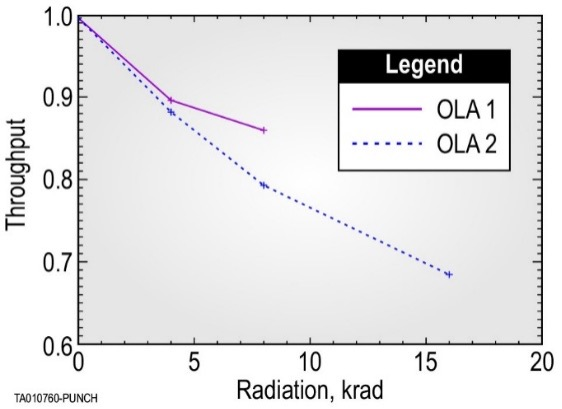}
    \caption{WFI prototype OLAs lost just 13\% to 21\% overall transmission after exposure to 8~krad of gamma rays, demonstrating resistance for the PUNCH 4~krad mission total dose.}
    \label{fig:radiation}
\end{figure}

Several of the OLA interior lenses are made of high dispersion rare-earth glasses, which are not radiation 
hardened and could, in principle, limit sensitivity late in the mission via radiation darkening of the glass. To 
demonstrate longevity, we exposed two prototype OLAs to penetrating gamma radiation to 8 and 16~krad.  These
values represent a Radiation Dose Margin of 2 and 4, respectively, against the estimated 2-year dose of 4~krad 
in PUNCH's polar orbit.  Figure \ref{fig:radiation} shows overall throughput response. At 8 krad the overall 
transmissions were measured to 79\% and 87\%, respectively, of the original on-band transmission 
(Figure \ref{fig:passband}). This is well within the sensitivity margin for the instrument as a whole, and is also
conservative: the dominant radiation source in PUNCH's orbit is low-energy particles with less penetrating power than 
the gamma rays used for dose testing.  In flight, the leading element, which is rad hard BK7 glass, provides 
significant shielding to the less hardened high-dispersion elements.

\begin{figure}[b]
    \centering
    \includegraphics[width=0.6\linewidth]{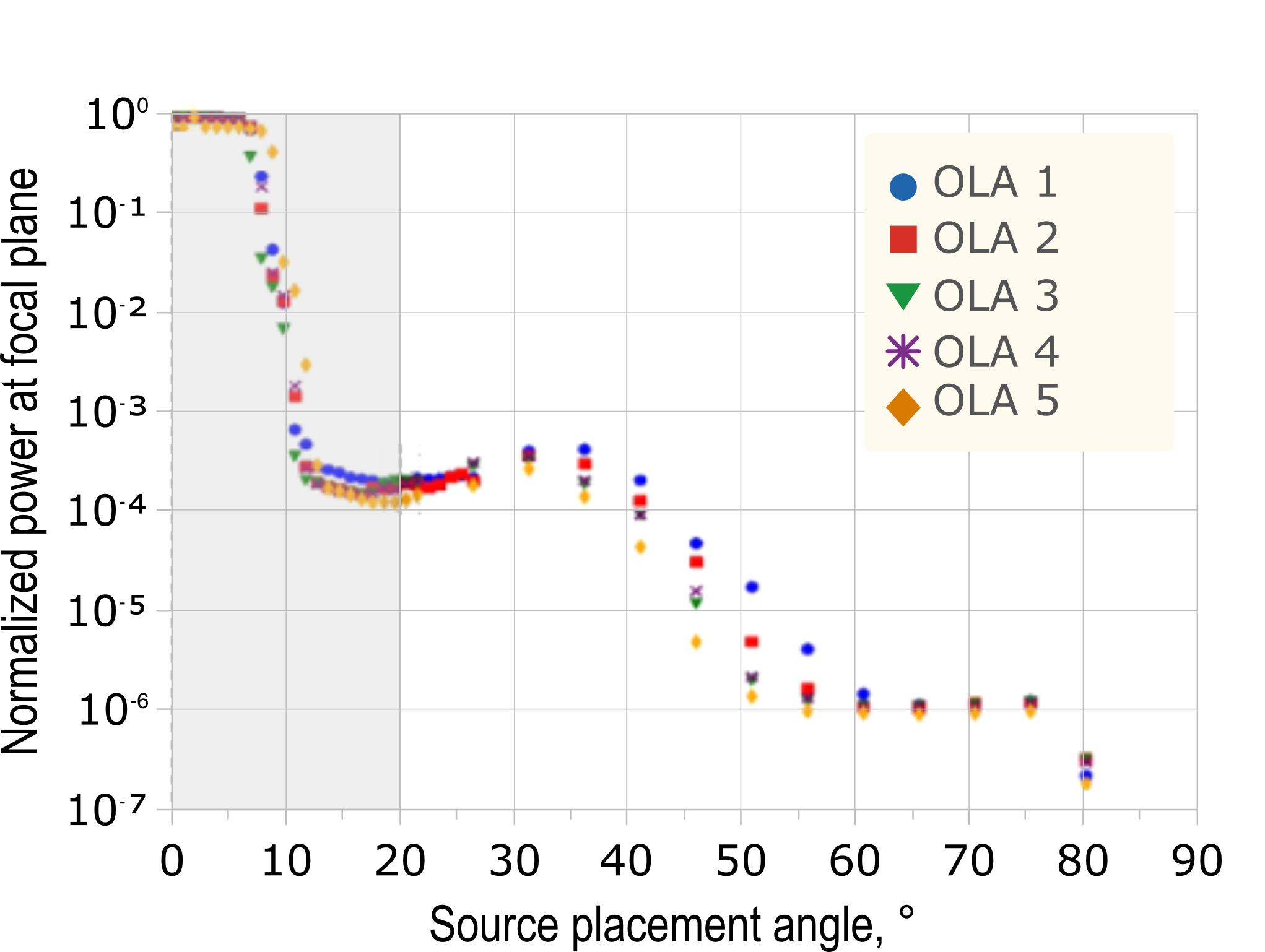}
    \caption{Internal stray-light results from five flight-qualified OLAs show strong rejection of stray light. Each OLA was focused on a source subtending 16$^\circ$ of angle near the center of the image plane. The active field of view is marked with a gray box. Wide-angle scatter of this broad source is better than $2\times 10^{-4}$.}
    \label{fig:OLA-stray-light}
\end{figure}

We tested the OLA optics for assembly-level stray light performance by focusing the image on a ground-test camera and mounting the OLA-camera assembly on a rotation stage near a single bright
circular light source subtending 16$^\circ$ in diameter.  We measured brightness near the center
of the field of view, as a function of rotation angle of the stage, by averaging pixel value over
the central portion of the image and adjusting exposure time to match the dynamics of the image.
Performance was better than 
$2\times 10^{-4}$ with this source, which subtended approximately 320,000 times the PSF size in 
solid angle.  This
measurement is important for characterizing overall stray light in the instrument, which is 
discussed in Section \ref{SS-testing}. Results are shown in Figure \ref{fig:OLA-stray-light}.

\subsection{Camera Assembly}\label{SS-Camera}

\begin{figure}
    \centering
    \includegraphics[height=0.25\linewidth]{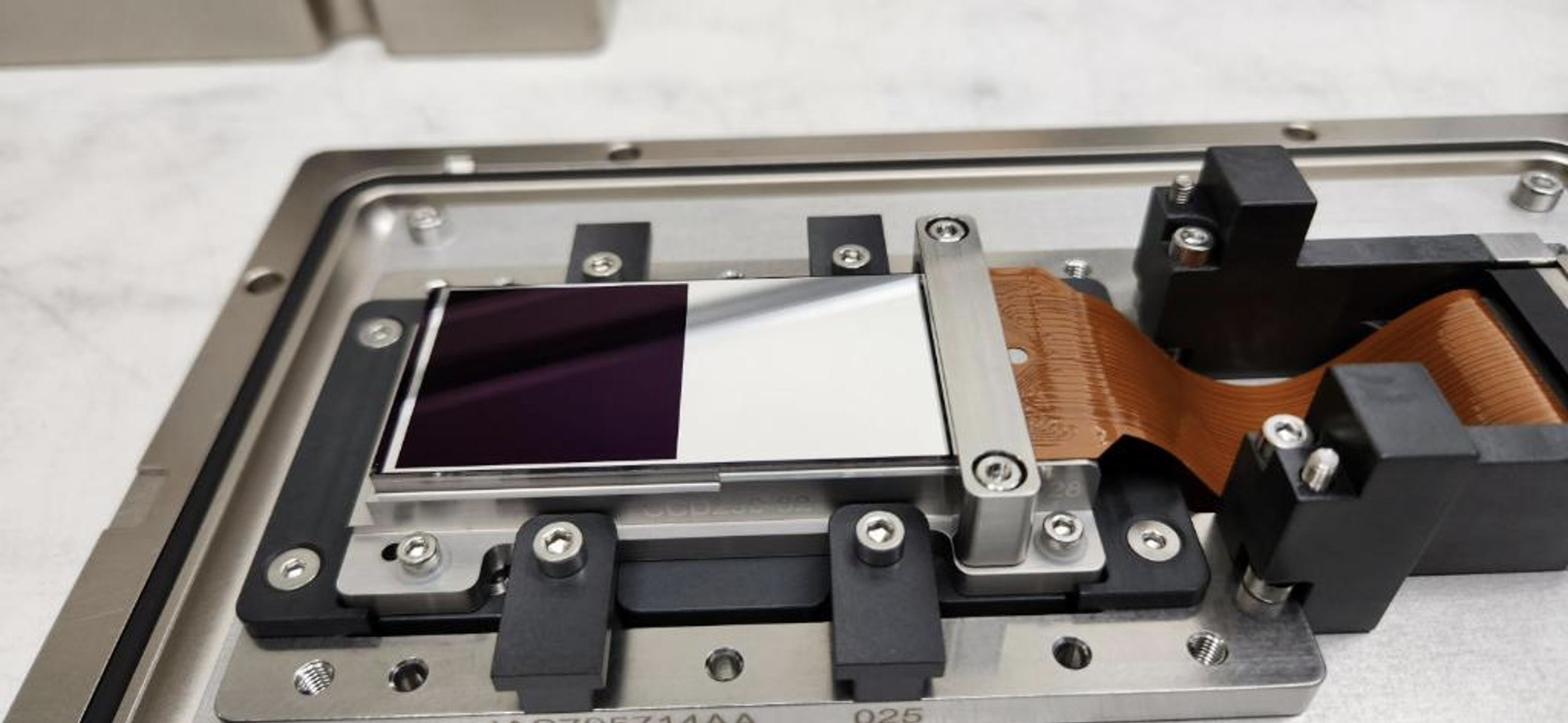}
    \includegraphics[height=0.25\linewidth]{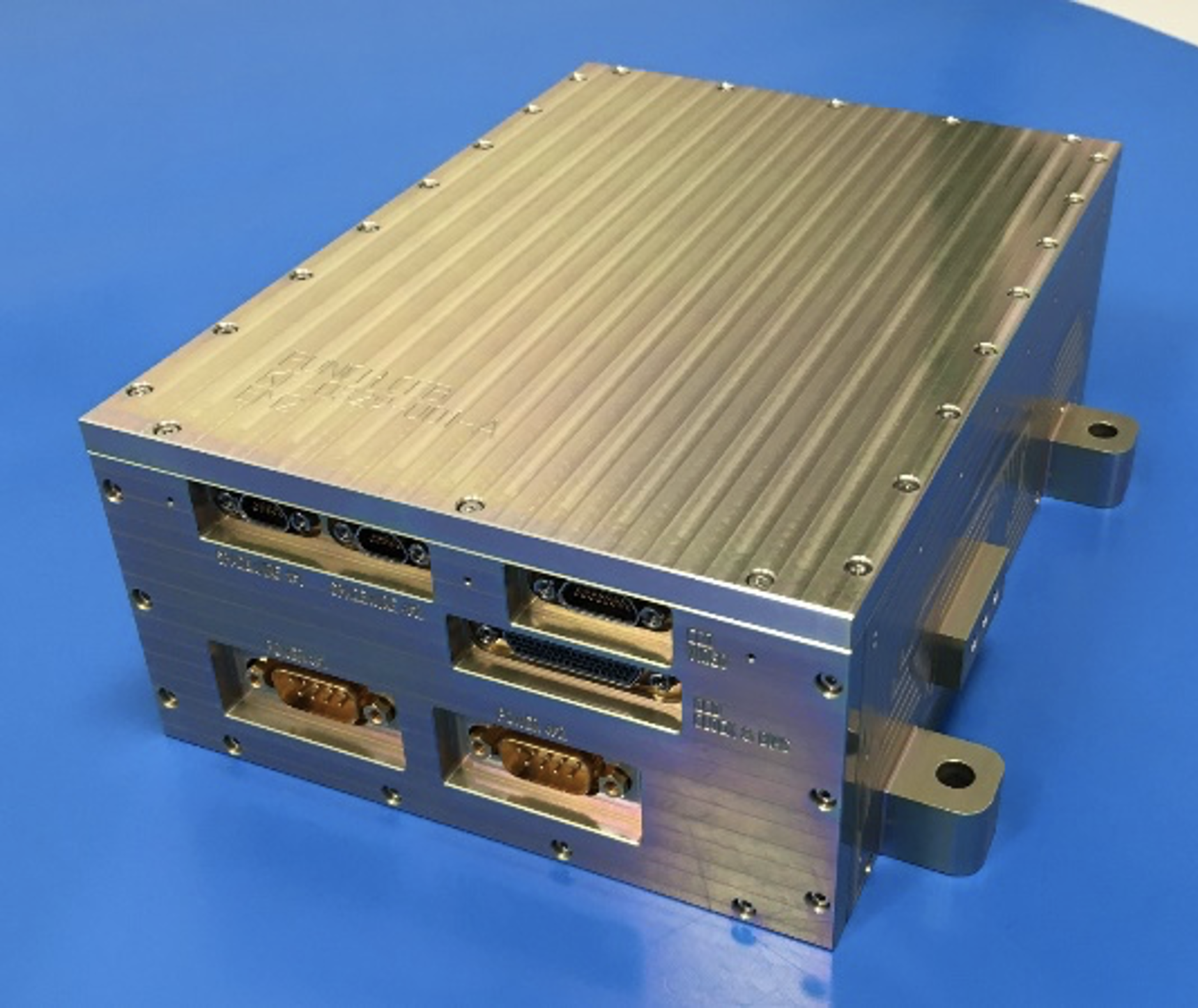}
    \caption{The PUNCH detector (left) is a Teledyne E2V CCD230-82 chip, operated in frame transfer mode. It is controlled and read out by camera electronics made by RAL Space (right).}
    \label{fig:CCD}
\end{figure}

Common to all four primary PUNCH instruments (Each WFI and NFI) is a charge-coupled 
device (CCD) camera system. The sensor is a Teledyne e2v back-illuminated, 
thinned CCD230-82 chip (Figure \ref{fig:CCD}). This is a 15~$\mu$m pixel device in their standardized CCD230 line, selected for good dark current, robust linearity, 
high quantum efficiency, fair radiation resistance, and capacity for frame-transfer
operation.  It is operated in frame transfer mode for stability of 
exposure time and to eliminate a 
mechanical shutter; this mode dedicates one 2k$\times$2k
area for image acquisition and 
another, coated with a metallic shield, for storage during readout. CCD accumulated 
charge values are digitized to 16 bits per 
pixel via dual-channel readout and control electronics made by RAL Space. Typical full-well depth 
for this device is tested to be over 500~ke$^-$ per pixel\footnote{We note that recent Teledyne 
E2V data sheets for the 230 series quote a much lower full-well depth of 150~ke$^-$.; this is 
because their default configuration is now quoted as the ``AIMO'' operating mode, which reduces 
dark current substantially at the cost of also reducing effective well depth. PUNCH does not operate in the AIMO mode.}; the PUNCH 
digitizers operate at 16 bits and a gain of
4.92 e$^-$/DN, for an effective well depth of 322k $e^-$.  Using effectively half of the CCD
capacity enhances linearity while preserving sufficient dynamic range for the WFI target of
deep space interspersed with occasional bright stars.    The detector surface is coated 
with the Teledyne-e2v
``midband'' catalog coating. The measured quantum efficiency is above 95\%, averaged across the 
450--750~nm passband of the optics. 

On-orbit, the nonlinear flat field is maintained by exposure scans with stimulation LEDs that are 
pulse-width modulated using an ultrastable, switchable current supply. This strategy provides 
rigorous maintenance on-orbit of the photometric calibration, provided only that the detector is 
reciprocal (yields values dependent on total fluence and independent of temporal pattern within 
an exposure).  On-orbit measurements with the LEDs showed that the WFI CCDs are all linear to
better than 0.1\%
even before nonlinear flat field correction.

The CCD is mounted on an Invar registration carrier that also serves as a heat sink to 
maintain the CCD operating temperature.  Electrical connection is via a short flexible 
cable that was bonded directly to the CCD substrate by Teledyne e2v.  The flexible cable
attaches to an interconnect module (ICM) headerboard from RAL Space that conditions 
and amplifies the analog
signal from the CCD,
and interfaces via a proprietary connector and shielded cable to a separate
Camera Electronics Box (CEB) that is mounted behind the instrument on top of the PUNCH
spacecraft. The CEB drives the CCD voltages and clock signals, and digitizes the ``video'' analog signal from the chip.  It interfaces to the host spacecraft via 
SpaceWire \citep{SpWr-standard}.  

The CCD nominal operating temperature is -65~C; it is cooled by a thermal strap 
connected to a shadowed radiator in the rear of the instrument (Section \ref{S-thermal}). 
At this temperature, read noise is 
less than 5 DN and the noise level is dominated by photon-counting ``shot'' noise across most
of the dynamic range.

A PUNCH exposure cycle consists of a series of clear/reset commands to fully clear the latent
image of the CCD, followed by an exposure interval, then a frame transfer operation that shifts
the latent image from the collection area to the storage area of the chip.  Finally, the latent 
image is read out. The nominal read rate is 1 MHz in each of two readout channels.  The frame 
transfer requires 125~ms, and image readout requires 2~s.  On-orbit exposure times vary between 3.5~s (for NFI clear exposures) and 51~s (for WFI polarized exposures), so that streaks from the
frame transfer operation vary between 0.25\% and 5\% of the 
total fluence from any one source in the image.  This is small enough to be 
correctable, but large enough to require explicit correction in the ground reduction pipeline.

\subsection{On-board data processing}\label{SS-data-processing}

After being acquired, WFI digital images are processed on-board to reduce downlink volume 
while preserving the signal.  

During nominal science operations, images are square-root coded, reducing the WFI dynamic
range from 16 bits to 10 bits.  This operation matches noise level (from photon 
statistics) to digital step size across the dynamic range, both reducing the 
number of bits per typical pixel and also greatly reducing the  amount of quantum 
noise that must be represented by the data. The square-root coded images are masked
to include only valid/required portions of the image plane: image areas outside the 
field of view are set to a constant value.  The square-root coded, masked images are
compressed using the lossless JPEG-LS algorithm, which achieves 4.8$\times$
compression on valid data and $>100\times$ compression on masked portions of the image.
The overall result is $>6\times$ compression of each image, with no effective loss
of dynamic range or image structure.  The images are reconstituted on the ground using Bayesian 
square-root inversion \citep{deforest_etal2022a} to preserve the expected photometric value.
Ground processing is summarized by \citet{DeForest_etal2026} and detailed by 
\citet{hughes_etal_2025}.

The cameras are capable of downlinking raw, overscan and other specialty data products
for calibration and diagnosis of the detector.

\section{Mechanical Design and Performance}\label{S-mechanical}

\begin{figure}
    \centering
    \includegraphics[width=0.67\linewidth]{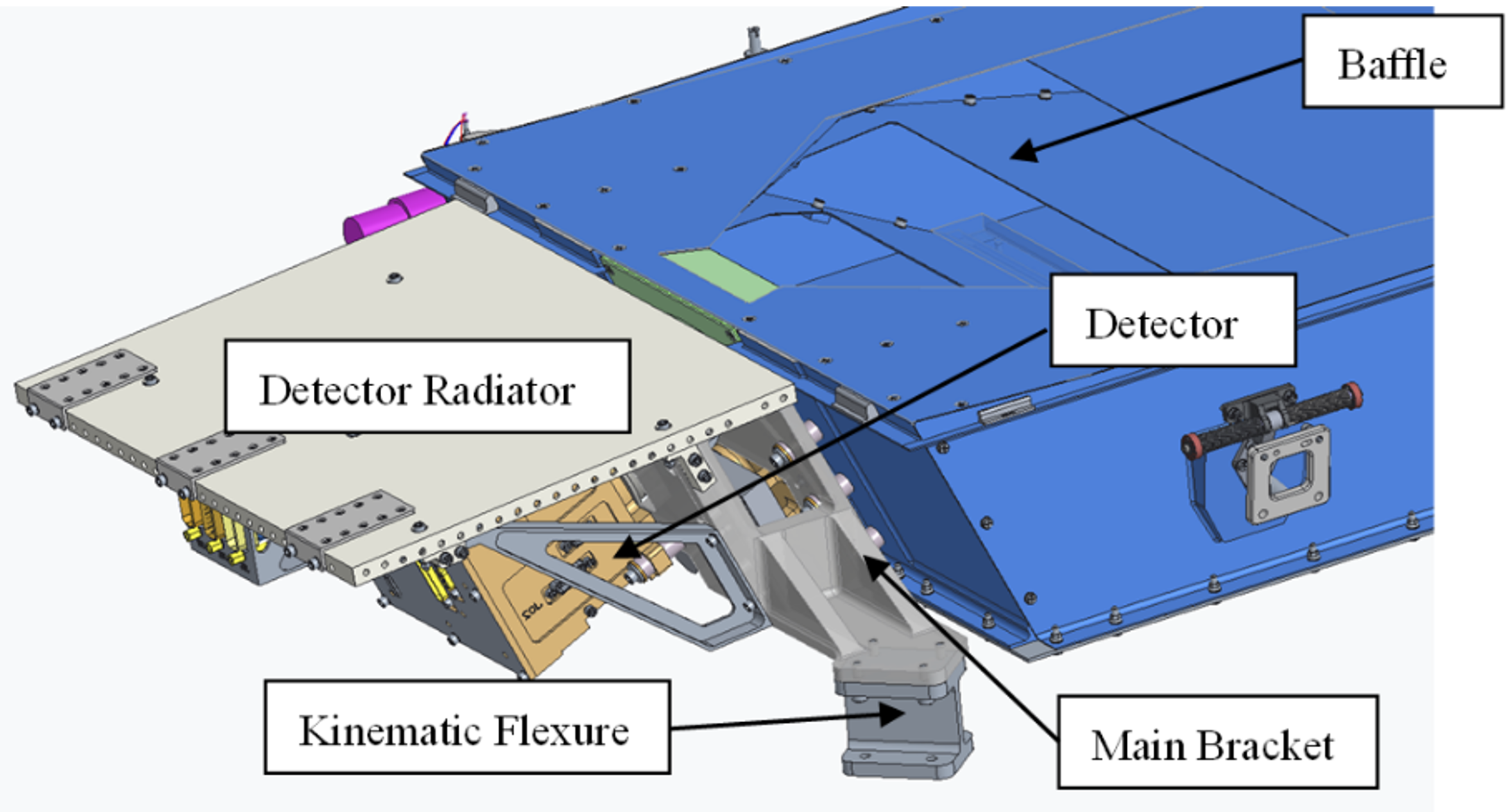}
    \caption{The WFI structure is built around a ``main bracket'' that is normal to the instrument boresight.}
    \label{fig:wfi-mainbracket}
\end{figure}

As a deeply baffled camera, WFI is built around a
``main bracket'' that serves as a metering structure and major support for all surrounding 
systems (Figure \ref{fig:wfi-mainbracket}). The main bracket is machined from a single 
block of aluminum to tighten machining/assembly tolerances.  
The main bracket is 
perpendicular to the optical boresight of the instrument, and is therefore pitched "up" (+Z) 
relative to the spacecraft XY plane (using the PUNCH spacecraft coordinate system, which is 
also shown in Figure \ref{fig:fov-for}). 
The main bracket supports the camera assembly, which in turn supports the OLA, 
ensuring registration and alignment of the optical train.  The PFW is supported directly
from the main bracket.  Baffle alignment tolerance relative to the main bracket is looser
than that of the optical components, the baffle is therefore bolted but not pinned to the main 
bracket.  The PFW and radiator are also supported directly by the main bracket (Figure 
\ref{fig:wfi-mainbracket}).  The instrument as a whole is supported by a three-flexure
type kinematic mount comprising three feet: two at the rear and one at the front of the 
instrument.  The flexures are fabricated from titanium, selected for its high strength and 
low thermal conductivity to isolate the instrument from the spacecraft.

During shipping, environmental testing, and launch, the WFI instruments are protected by
flat, lab-resettable, one-time-open doors that close out the internal volume, protecting 
the surfaces and 
structure of the baffles and the optics.  The doors are mounted on two spring-loaded
hinges on the -Y face of the instrument (the rear of which is visible in Figure
\ref{fig:wfi-mainbracket}), and latched closed with a spring-loaded ball-and-cup latch
assembly on the +Y face of the instrument.  In flight, the latch is actuated with a
high-output paraffin (HOP) release mechanism.  The door rotates more than 180$^\circ$ 
during actuation, so that it is entirely below the reference plane formed by the top
of the baffle. During launch, the doors are rigidly held only by the hinges (-Y) and 
two latch points (+Y), and drum modes of the door were therefore of interest during 
mechanical design and normal-mode analysis.

\begin{figure}
    \centering
    \includegraphics[width=0.8\linewidth]{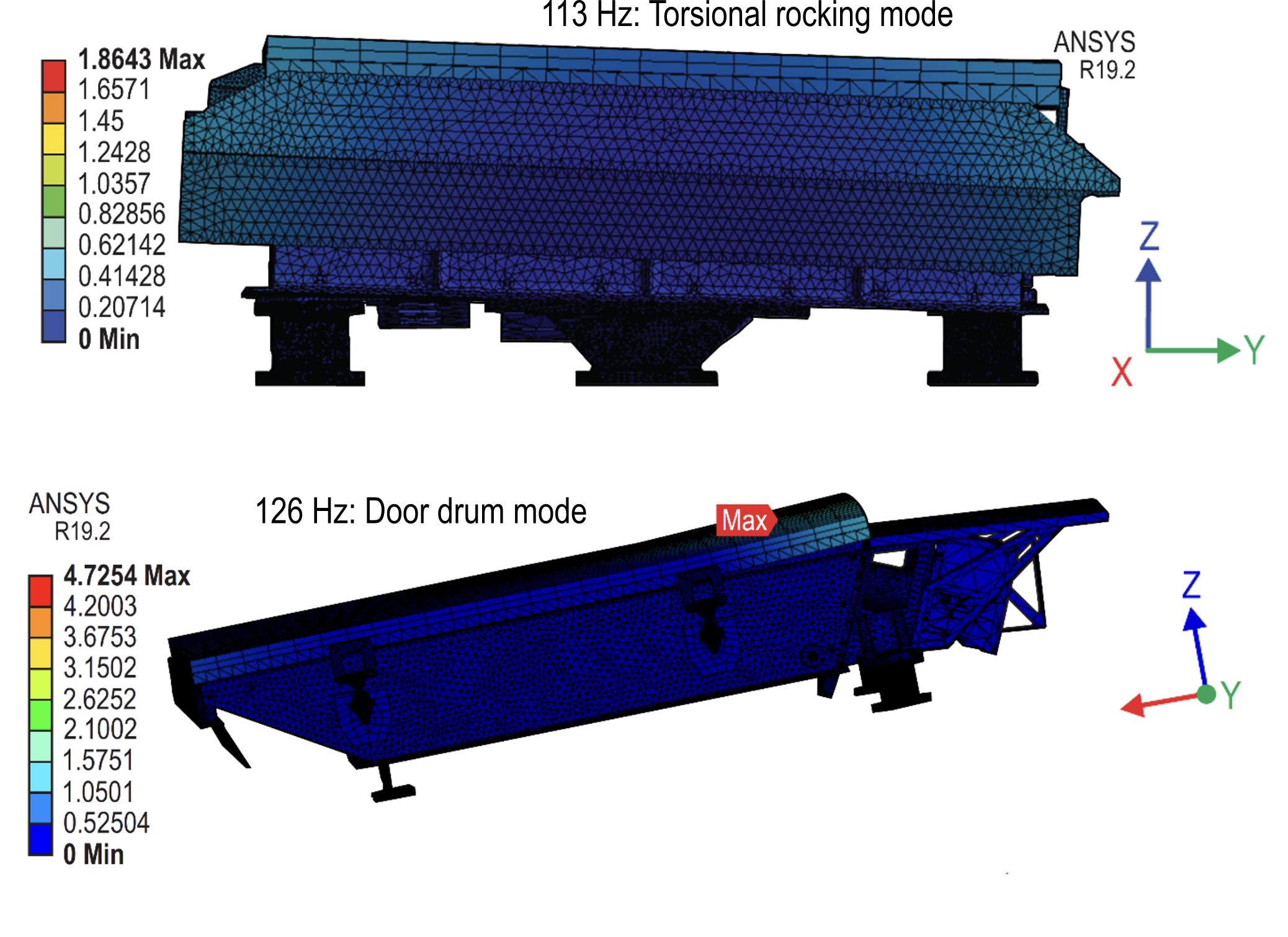}
    \caption{WFI instrument first significant normal modes (with more than 10\% mass 
    participating, on
    a normalized motion basis) are above 100 Hz, with predicted elastic motion under 1mm.  
    Top: lateral rocking, with body torsion applied; bottom: drum mode of the front of 
    the closed door.
    }
    \label{fig:wfi-modes}
\end{figure}

\begin{figure}
    \centering
    \includegraphics[width=0.8\linewidth]{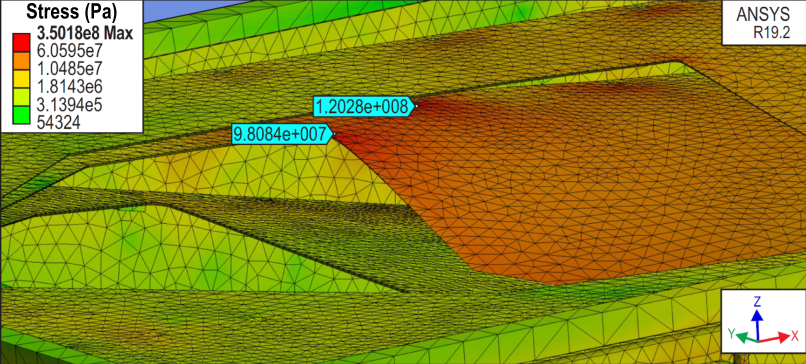}
    \caption{WFI fundamental mode is 37 Hz, has low mass participation, and is a flapping mode of the light trap vanes. Calculated maximum stress of 120 MPa is 3$\times$ below the yield threshold of the high-strength aluminum. All three WFIs' light trap baffles survived launch.}
    \label{fig:fundamental}
\end{figure}

A computational finite element model was used to predict the mechanism's normal modes, and stresses and strains within 
the WFI structure during launch.  All flight loads were considered, including sinusoidal, random
vibration and acoustic loading. The first ``significant'' mode (with more than 10\% of the
instrument mass participating) is a torsional rocking mode 
around the front foot, 
with frequency of 113 Hz. 
The second significant mode is a deformation of the closed door, at 126 Hz. These modes are 
shown in Figure \ref{fig:wfi-modes}. The large, flat plates of the light trap have lower 
fundamental modes but are light enough that they do not significantly affect the instrument as a
whole.  This fundamental mode, at 37 Hz, is shown in Figure \ref{fig:fundamental}.  The calculated 
maximum
internal stress, at 120~MPa, is a factor of 3 below the yield strength of the high-strength 
aluminum vanes.

 A protoflight test program was specified for all PUNCH hardware, including WFI. All three WFI 
 flight models were put through a full protoflight environmental test campaign, 
 including sine and random vibration structural loading.  After integration with the S/C, the 
 instruments were exposed to acoustic testing.  Testing confirmed WFI natural modes as predicted 
 by the finite element analyses.
 
\section{Thermal Design and Performance}\label{S-thermal}

\begin{figure}
    \centering
    \includegraphics[width=0.75\linewidth]{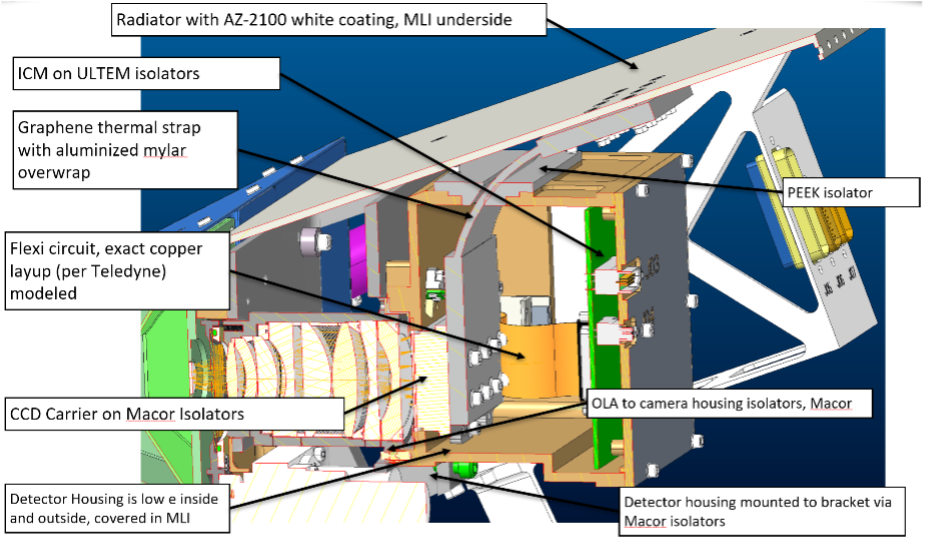}
    \caption{The WFI thermal design focuses on maintaining significant temperature differences inside the optics and camera assembly, and isolating the camera and instrument from the host spacecraft.}
    \label{fig:thermal-design}
\end{figure}

The WFI thermal design (Figure \ref{fig:thermal-design}) is driven by 
three major requirements: maintaining the focal plane 
CCD at -65°±0.5°C, keeping all components (OLA, filter wheel, electronics) within 
acceptable thermal limits, and isolating the instrument from the spacecraft. This 
requires managing significant thermal differences through the instrument
as the baffle system radiates to deep space, the CCD requires low temperature, 
and the electronics and optics require 
somewhat higher temperatures. The PUNCH 
operating mode maintains a nadir pointing side (-Z), a Sun-oriented side, (+X), and a 
dark side (-X), affording a stable thermal radiative environment. 
The design uses a cold-biased 
passive control system with both operational and non-op survival heater circuits. 
The instrument does not use active cooling. 

The CCD is cooled using an optimized passive radiator coated with thermal paint on the 
Zenith side and multi-layer insulation (MLI) on the opposite side.  
The coating material offers excellent long-term performance, strong infrared radiative 
properties, and minimal visible light absorptivity. A multi-layer graphene strap
is used to transfer heat from the CCD to the radiator. The radiator is sized to receive 
both CCD heat and parasitic loads while maintaining required CCD temperatures.  The 
design is cold biased, with trim heat from a simple thermostatically controlled on/off
electric heater on the CCD Invar carrier. The overall WFI design included accommodation 
to trim the CCD radiator with additional MLI as needed during instrument thermal 
balance testing.

The OLA and the camera housing are thermally isolated via three point mounted custom 
machined titanium washers, chosen for moderate thermal isolation and good dimensional 
stability.  Final thickness of the washers were determined during the OLA 
alignment process.  Macor (machinable glass-ceramic from Corning) was used to
structurally support and thermally isolate the camera housing from the main bracket. 
Cooling of the CCD with a dedicated radiator resulted in undesirable cooling of the 
interconnect module (ICM) by way of the flex cable connection between 
the two.  For this reason, the ICM was independently thermally controlled and isolated 
from the detector housing with ULTEM (amorphous thermoplastic polyetherimide) spacers.  The ICM remains a significant source of heat
flux into and through the CCD, and out the radiator, which was accounted for in the 
radiator sizing.

Most of the instrument is covered with 12-layer embossed MLI for exposed surfaces.  The 
working side of the WFI baffle was coated in either Aeroglaze Z307 black polyurethane 
paint, or 
Acktar Vacuum Black, to minimize scattered light feeding into the optical path.  This 
side is the Zenith facing side of the PUNCH observatory, and therefore drove the baffle to 
very cold temperatures, approaching -100°C.  To prevent pulling the camera end of the 
instrument down to undesirable temperatures, thermal isolation between the baffle and the 
main bracket is provided by Macor spacers.  Added heat to the baffle was not an option 
for overall power budget reasons. Baffle positional tolerances were broad enough that
uniform thermal contraction of the baffle as a whole was acceptable.  

The design phase thermal models correlated very well with actual thermal vacuum testing.  
Table \ref{tab:thermal-correlation} shows good agreement between pre-integration 
Thermal\-Desk\-top models and thermal balance testing in vacuo.

{\centering
\begin{table}[]
    \centering
    \begin{tabular}{c|c|c|c}
       System & Test (C) & Prediction (C)  & Difference (C)\\
        \hline     
        Main Bracket & -52 & -52 & 0\\
        Cold Finger & -76.4 & -74 & -2.4\\
        ICM & -22 & -18 & -4\\
        Radiator & -81 & -79 & -2 \\
        OLA & -32 & -28 & -4\\
        PFW & -19 & -10 & -9\\
        PFW motor & -24 & -13 & -11\\
        Baffle right & -103 & -108 & 5 \\
        Solar Vane & -106 & -109 & 4 \\
        Shroud & -147 & -147 & 0    
    \end{tabular}
    \caption{WFI thermal vacuum tests showed good agreement with pre-integration modeling.   }
    \label{tab:thermal-correlation}
\end{table}
}

\section{End-to-end Performance Testing} \label{S-testing}

Each WFI instrument underwent end-to-end testing at three key stages: 
immediately after final assembly, after thermal vacuum testing, and after vibrational 
testing. During each of these end-to-end tests, the following tests were performed: 
optical focus,  optical vignetting, orientation of each polarizer in the PFW, 
performance of the in-camera LEDs (which included as a bonus a measurement of the size 
distribution of CCD surface contaminants), and functionality of the on-board
heaters. The end-to-end tests were performed at room temperature and 
pressures in an ISO Class 7 cleanroom, with the WFI instrument mounted on a Flotron 
rotation stage, allowing pitch (rotation about the Y axis) of the instrument 
to be controlled within $\pm 0.1^\circ$.  End-to-end stray light testing was 
executed between final assembly and environmentals of WFI-1, and based on the
results, which showed ample margin on stray light performance, no further end-to-end
stray light tests were carried out.

\subsection{Stray Light testing}\label{SS-testing}

The WFI-1 instrument was tested end-to-end for stray light performance, at NRL's Solar 
Coronagraph Test Chamber (SCOTCH) facility \citep{korendyke_1993,Morrill2006}.  SCOTCH 
is a dedicated vacuum chamber facility, designed to eliminate
atmospheric scattering from coronagraph stray light measurements. It includes a 
high-voltage arc lamp and low-scatter feed optics to illuminate the instrument under
test with collimated light from 
either an approximate point source of a simulated Sun.  A long baffled tube separates
the light source from the instrument under test.

\begin{figure}
    \centering
    \includegraphics[width=0.6\linewidth]{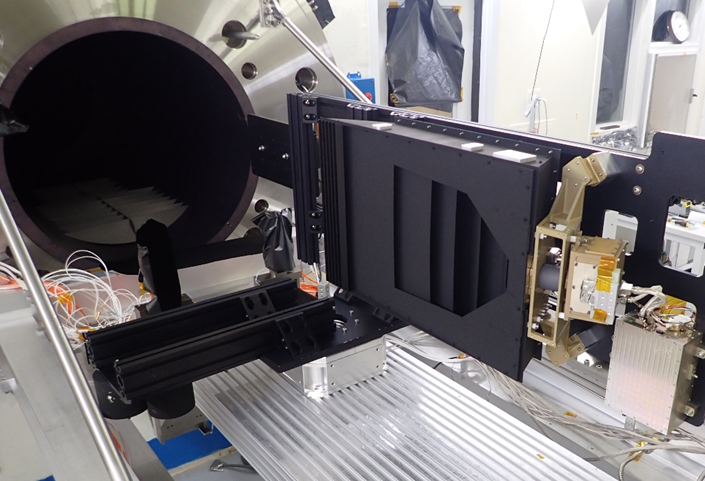}
    \caption{WFI-1 was stray light tested at NRL's SCOTCH facility \citep{korendyke_1993}; it is shown here mounted on a metering structure and rotation stage, ready to be sealed in the vacuum chamber.}
    \label{fig:scotchg}
\end{figure}

WFI-1, supported on a purpose-built black-anodized metering structure, was mounted on a 
rotational stage in the SCOTCH test chamber (Figure \ref{fig:scotchg}), allowing 
for active control of pointing in the pitch direction during test.  Rather than cool
the chamber walls to simulate deep space, the WFI-1 CCD was cooled with a copper cold strap
linked to a facility-provided cold plate cooled with liquid nitrogen. An adjustable slit
mounted at the entrance aperture of SCOTCH controlled the angular extent of the simulated
solar beam.

\begin{figure}
    \centering
    \includegraphics[width=0.8\linewidth]{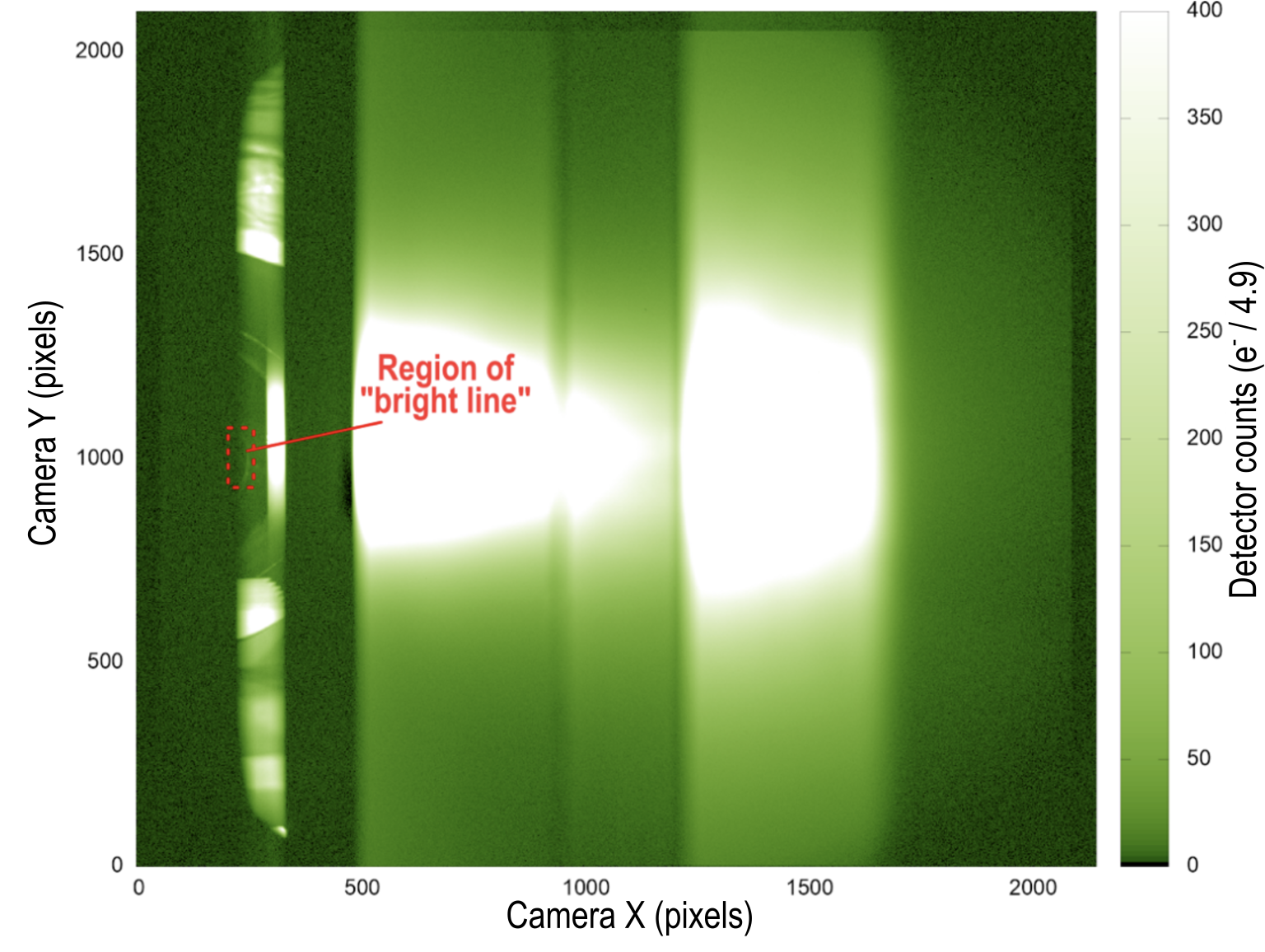}
    \caption{Image acquired in the ``low'' beam condition (see text) through WFI-1
    of the SCOTCH interior with a 10 second exposure time 
    shows chamber stray light to be much brighter than Fresnel diffraction around the solar baffle.  The baffle is the dark 
    structure at far left (X=0 to X=200). The SCOTCH baffled vacuum tunnel is visible from X=250 to X=350.  The stray light
    reducing cowl is visible between X=350 and the right-hand side of the image. The cowl is 
    illuminated by backscatter from the beam trap at the rear of the chamber.  The vacuum tunnel is illuminated by backscatter
    from the front of V1. Both scattered sources are much brighter than the ``bright line'' (marked) where the simulated solar 
    beam crosses the solar baffle.
    }
    \label{fig:scotch-image}
\end{figure}

With the incoming beam directed at the knife edge of the first solar vane (V1) at the front of
the baffle, the WFI instrument was rotated 0.6$^\circ$ (pitch up) relative to the light source
to simulate the in-flight position of the Sun relative to the instrument XY reference plane. The
beam was clipped to 5$\times$83~mm$^2$ cross section, with the narrow (5~mm) direction perpendicular to, and the
long (83~mm) direction parallel to, the edge of V1.  The beam was adjusted to clip the V1 knife edge, with 2.5~mm of the beam 
spilling 
over the WFI instrument to a conical light trap at the back of the chamber.  
A removable light shroud (cowling) surrounded
the instrument to further reduce specular reflections of the beam off the chamber wall back 
into the WFI optics.  Figure \ref{fig:scotch-image} shows the view, through WFI, of the 
shroud and the SCOTCH interior. Most of the field of view is dominated by the stray light 
reducing cowl, which occupies nearly all lines of sight to the right of X=350~pixels on the image.  In 
the image, it comprises five visible planar surfaces, all of which are out of focus due to proximity. The leftmost plane, extending between X=350 and X=500, is seen at a glancing angle and appears dark. The narrow band between X=200 and X-350 comprises lines of sight that extend into the vacuum tunnel toward the light source. The dark feature to the left of X=200 is the solar baffle. The marked region is the locus where Fresnel diffracted light is expected. At this brightness scale, the image is dominated by the stray light in the tunnel and on the cowl.  The tunnel stray light arises mainly from beam scatter off of V1, and the cowling stray light arises mainly from beam scatter off of the beam dump at the rear of the chamber.

We used a differential measurement to minimize the effects of chamber stray light 
and increase sensitivity.
The light source was adjusted at the
slit, to 
direct the 5$\times$83~mm beam
between three configurations: ``hit'' (full beam blocked by V1), 
``low'' (beam bisected by V1), and ``high'' (full beam passing over V1
without interacting directly).
The ``hit'' and ``high'' measurements characterized two major patterns of scattered light in the 
chamber, with the 
``low'' configuration being a linear combination of ``hit'', ``high'', and the desired 
measurement of 
Fresnel scattering around the solar baffle:
\begin{equation}
    B_{scatter} = B_{low} - \left(\alpha B_{hit} + (1-\alpha) B_{high}\right)/2\,,
    \label{eq:scotch}
\end{equation}
where each $B$ represents an image pixel value, $B_{scatter}$ is the intended measurement, the 
other three $B$ values are taken from corresponding beam configurations, and $\alpha$ is an 
adjustable coefficient near 0.5, which accounts for the fraction of the beam above and below the 
edge of V1 in the ``low'' case. We used that measurement method to overcome reflected stray light 
within the SCOTCH chamber.  It is based on the observation that the roughly 6~mm beam motion 
between the three cases is negligible compared to geometry other than the V1 edge passage.

The excess pixel value calculated using Equation \ref{eq:scotch} and $\alpha=0.5$ was under 4 DN
in this 10 second exposure.  Folding the total intensity (summed brightness) of the Fresnel bright
line through the scattering profile in Figure \ref{fig:OLA-stray-light} yields an estimate of 
overall solar stray light attenuation at the center of the field of view of better than 
$10^{-16}$ for WFI-1, more than two orders of magnitude better than the instrument requirement of
$10^{-14}$.  This result was borne out in flight and is further discussed in Section 
\ref{S-discussion}.

\subsection{Focus testing}\label{SS-focus}

To measure WFI’s optical focus, we constructed an artificial “star” from a 
stabilized broadband light source, a fiber optic cable, and a pinhole. This 
star was set up beyond the 42-foot hyperfocal distance of the WFI camera subsystem (70 
feet was typical). The star was kept in a fixed position and the WFI instrument was 
rotated in both pitch and yaw such that the light from the star fell on different 
regions of the detector. Within the 2k$\times$2k pixel detector, we defined a 
5$\times$5 grid 
of non-adjacent 100$\times$100 pixel boxes.  We acquired an image of the star at 
each of the 25 
positions over the field-of-view, with the PFW in the clear (unpolarized) position.   
We captured additional polarized images at select positions at the center and edge of 
the field-of-view to verify consistency with the clear results. This Fourier / 
modulation-transfer function method matches well the way WFI data are prepared in the 
flight ground system: every flight image is normalized to a uniform PSF using the methods of 
\citet{hughes_2023}, so that Fourier component attenuation is the most relevant
measure of focus.  

\begin{figure}
    \centering
    \includegraphics[width=0.45\linewidth]{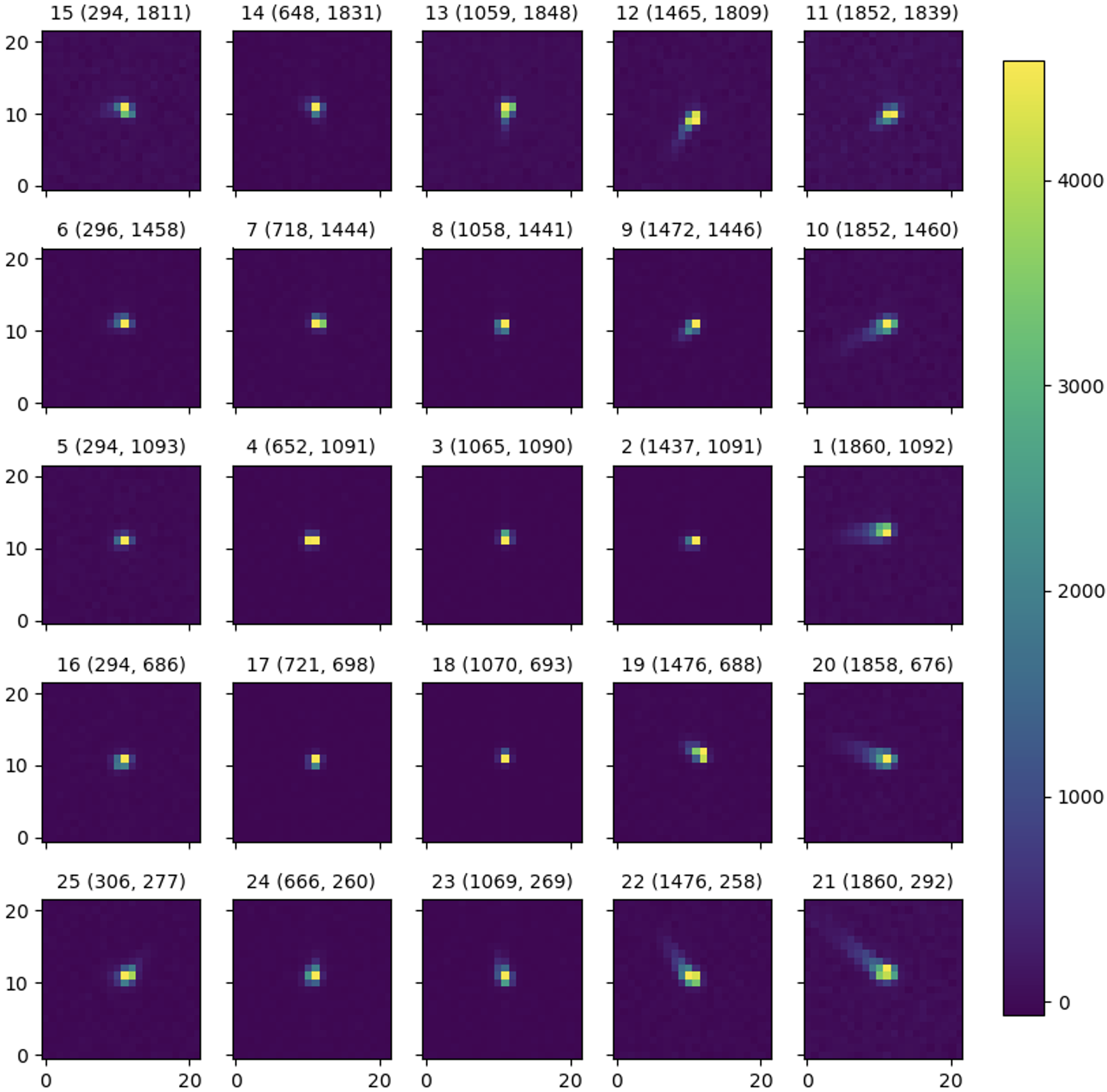}
    \includegraphics[width=0.45\linewidth]{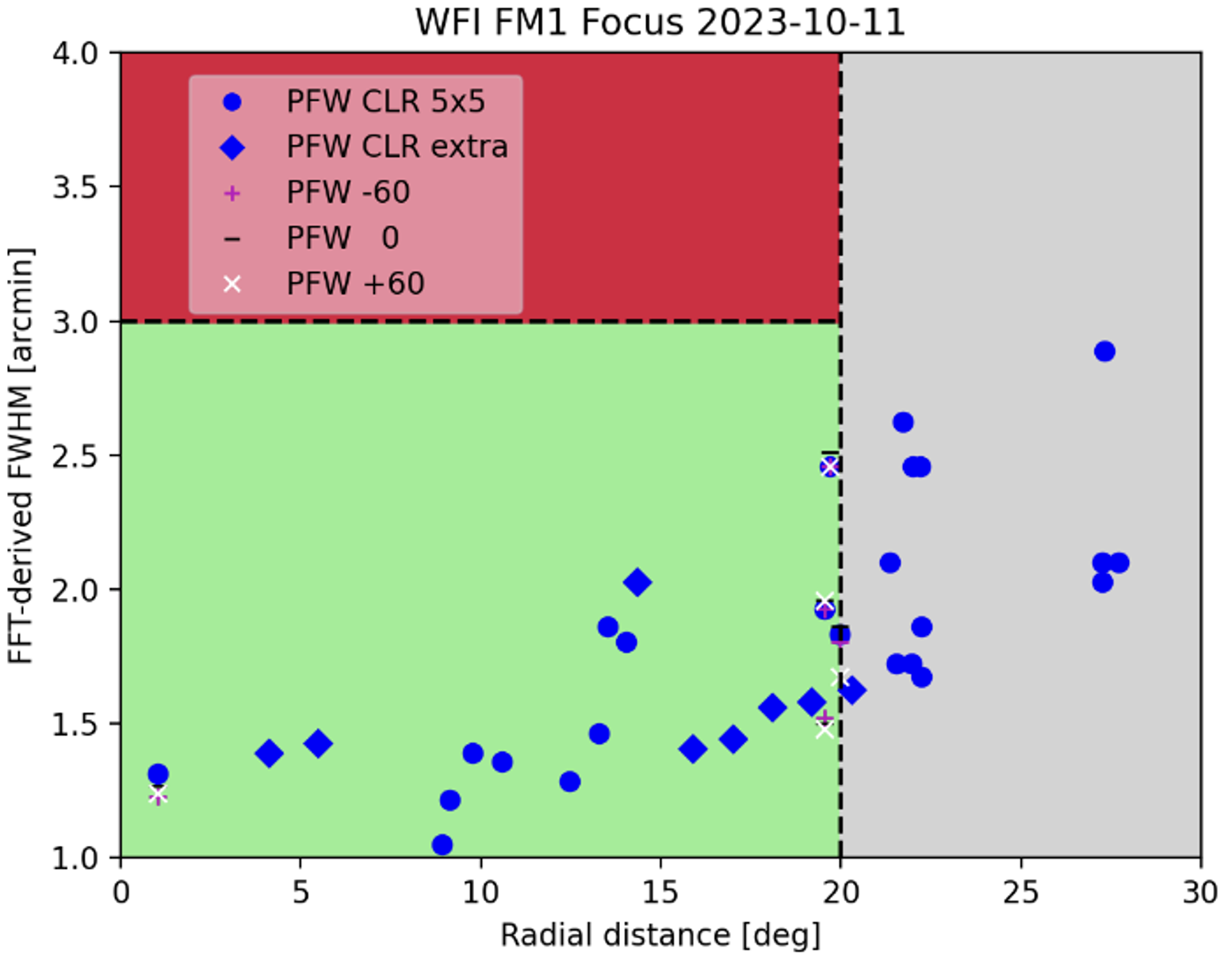}
    \caption{Sample focus-test results from WFI-1 show methodology. Left: WFI-1 focus testing PSFs sampled the detector at 25 representative locations. Right: FFT-derived equivalent spot widths, as a function of radial distance from the center of the detector, verify focus quality.  Points in the green region meet the requirement of 3 arcminutes.  Points in the grey region meet requirements despite being outside of the required field-of-view.}
    \label{fig:focus}
\end{figure}

We measured the artificial star's focus at each grid position. Because of 
substantial coma near the edges of the field of view, we found a simple Gaussian PSF 
fitter to be insufficient: measured residuals compared to a Gaussian PSF were highly 
structured, and the measured parameters depended strongly on the star’s phasing position 
relative to the pixel grid. Instead, we computed the 2-dimensional FFT of a small image 
patch centered on the star, computed the total power as a function of spatial frequency 
$k$ (neglecting the 0-frequency term), fit a cubic spline to those 
points, found the frequency at which the response was half of the peak response, and 
converted that frequency to a spatial distance. This measurement approach resulted in a 
stable and repeatable FFT-derived full-width at half maximum (FWHM), which we used to 
verify the focus requirement.  An example result is given in Figure \ref{fig:focus} for WFI-1.  All three WFI instruments met their focus requirements with margin and 
showed stability
of focus across environmentals.

\subsection{Vignetting function testing}\label{SS-vignetting}

To measure the vignetting caused by the optical system and by the solar vanes, we used 
the same Flotron bench mount as for the focus test, replacing the artificial star with a 
flat white projector screen.  Achieving a flat illumination over the entire WFI field 
of view proved challenging for several reasons: the screen separation needed to be far 
enough from the instrument to prevent electrostatic discharge; the ideal surface would 
require a radius of curvature equal to the distance from the instrument; the screen must 
be uniformly illuminated but not saturate the detector.  In practice, the screen was not 
large enough to fill the WFI FOV in a single exposure, the screen was flat and not 
curved, and low-level diffuse room lights were found through trial and error.  The
ground-derived vignetting functions were sufficient to verify requirements, but were
supplemented with in-flight end-to-end testing using the starfield for detailed 
calibration.

We pitched WFI on the Flotron such that the screen filled slightly more than 
half the field of view, acquired images of the screen, then adjusted pitch
so that the screen filled slightly more than the other half of the field of view. In 
post-test analysis, these two half-flat-field images were digitally spliced along the 
natural boundary of the two halves of the detector, creating a composite flat-field 
image. No effort was made to correct for different bias levels of the two halves of the 
detector or for differences in lighting across the projector screen. In analysis, the 
left half of the image was used to measure the vignetting due to the solar vane, and the 
right half of the image was used to measure the vignetting due to the OLA. 

\begin{figure}
    \centering
    \includegraphics[width=0.5\linewidth]{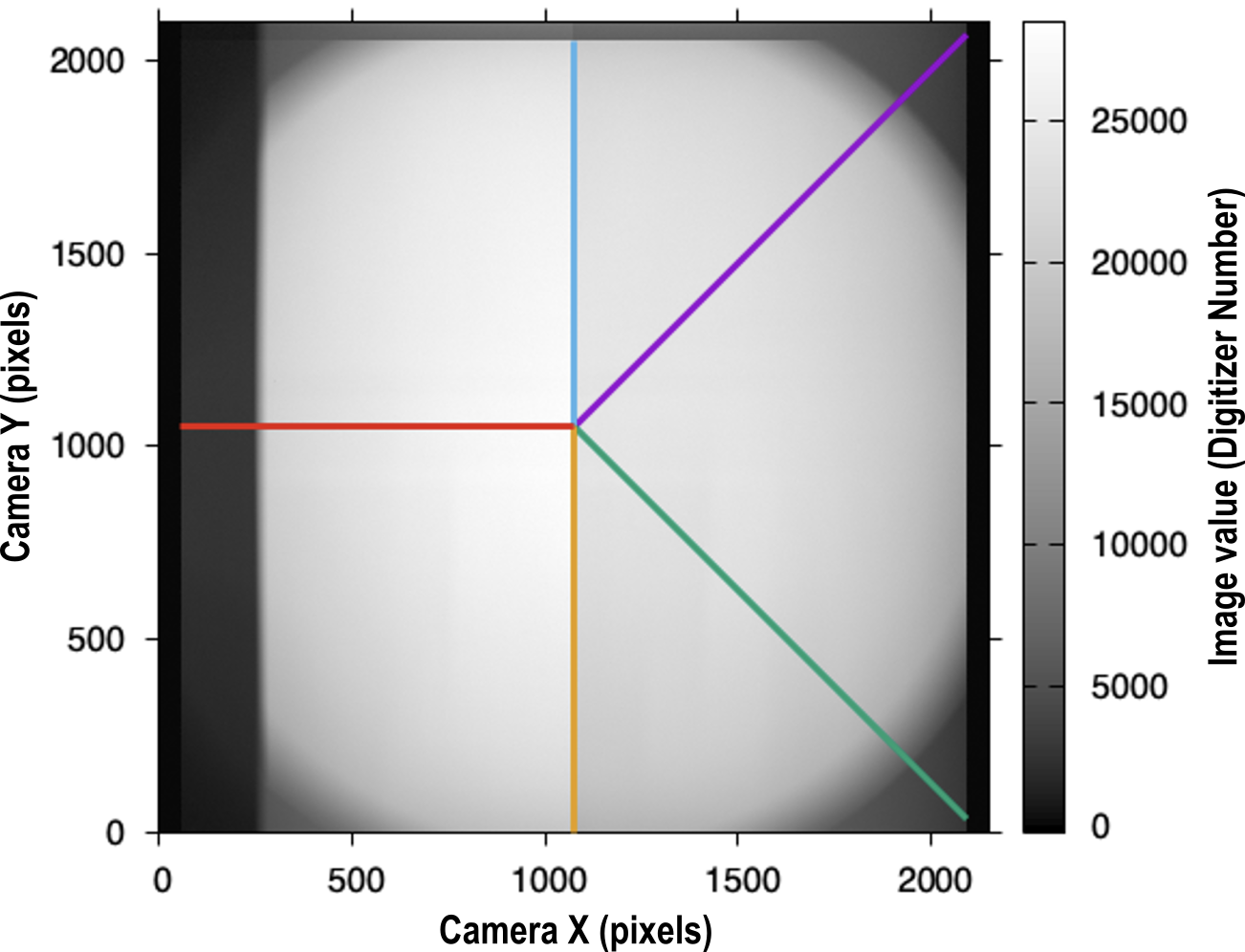}
    \caption{Composite camera image from a WFI-1 vignetting test shows major features under test: central division from the dual readout; solar vane at left; and 
    far-field vignetting at right. Each line corresponds to a trace in Figure \ref{fig:vignetting-plots}.}
    \label{fig:vignetting-image}
\end{figure}

\begin{figure}
    \centering
    \includegraphics[width=0.9\linewidth]{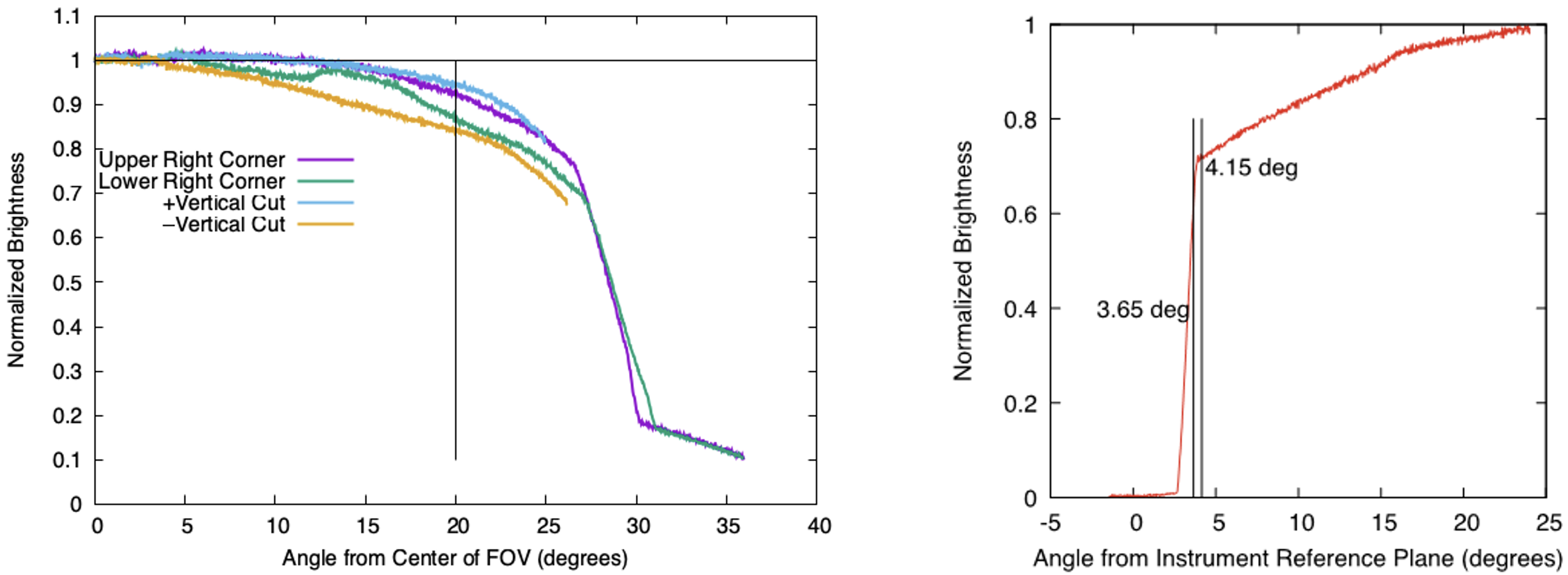}
    \caption{Brightness profile cuts of WFI-3 flatfield image in Figure \ref{fig:vignetting-plots}, as a function of radial distance from image center and normalized to the image center intensity, show quantitative vignetting performance that meets requirements.  Left: Outer field-of-view vignetting profile demonstrates that the OLA cutoff meets the requirement of $>50\%$ at 20 degrees. Right: Inner field-of-view vignetting profile demonstrates that the solar-vane cutoff meets FOV requirements of $>0$ at 3.5$^\circ$ and $>80\%$ at 5$^\circ$.}
    \label{fig:vignetting-plots}
\end{figure}

A representative flatfield and corresponding intensity profiles for WFI-3 are shown in 
Figures \ref{fig:vignetting-image} and \ref{fig:vignetting-plots}.  The far-field 
vignetting functions were in general found to match well the geometric vignetting from 
projection of the aperture into angles off the boresight.  The near-field vignetting 
functions followed closely the geometric form from a straight line (the solar baffle 
edge) intersecting the circular optical aperture.   Both functions were found to meet
requirements and to be 
stable across environmentals for all three WFIs.

\subsection{Polarizing Filter Wheel alignment testing}\label{SS-pfw}

We verified that the orientation of the three linear polarizers in each PFW met 
alignment requirements an were maintained at $\pm$~$60^\circ$ relative to one another. 
For this test, we fed light from a stabilized broadband source through a fiber optic 
cable into an integrating sphere, thence through a linear polarizer mounted in a 
precision rotating mount. We used a digital microscope to measure accurately the angle 
of this laboratory linear polarizer. We pointed WFI at the laboratory polarizer and 
acquired images at each of the three PFW positions, over a range of linear polarizer
angles in 1$^\circ$ increments. We performed aperture photometry on the collected 
images, to
measure variation of spot brightness with laboratory polarizer angle.

\begin{figure}
    \centering
    \includegraphics[width=0.9\linewidth]{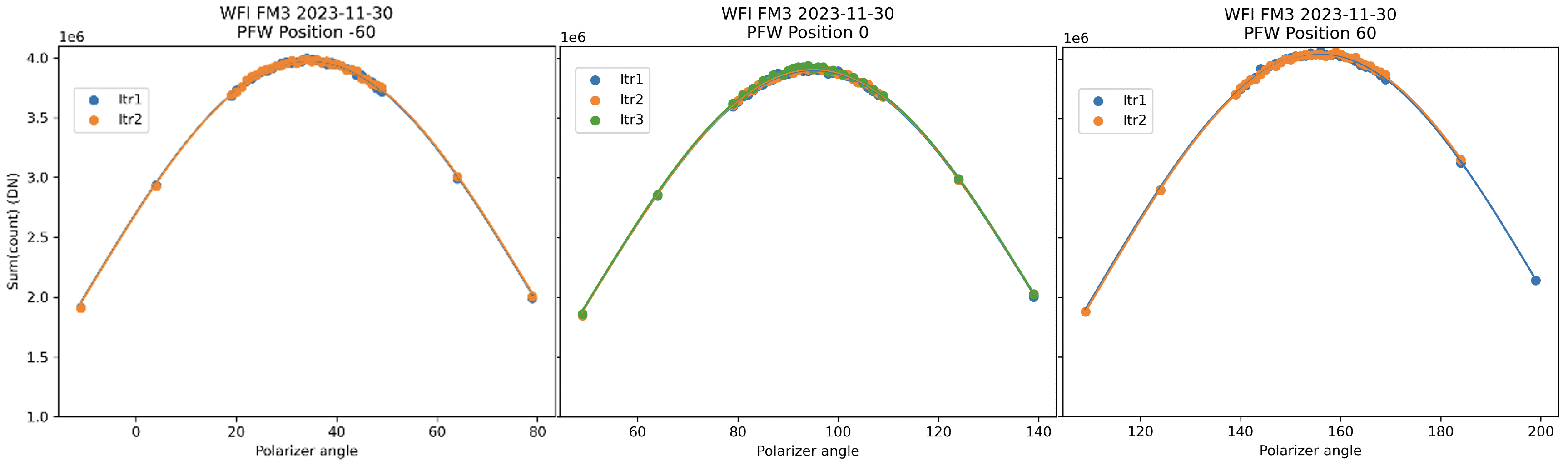}
    \caption{Intensity profiles as a function of GSE linear polarizer angle for each of the three WFI-3 PFW positions reveal correct alignment of the three polarizers in the WFI-3 PFW.}
    \label{fig:pfw-alignment}
\end{figure}

The light counts versus angle for two iterations of the same PFW position are shown in 
Figure \ref{fig:pfw-alignment}. We fit a sine wave to the brightness profile, to 
determine the central brightest angle $\theta_i$ for each PFW position (with $i$ varying across \{M,Z,P\}). Multiple 
iterations per position yielded consistent results, and the alignment was found to be well within the 1$^\circ$ relative angular tolerance.

\section{Discussion \& Conclusion}\label{S-discussion}

\begin{figure}
    \centering
    \includegraphics[width=0.9\linewidth]{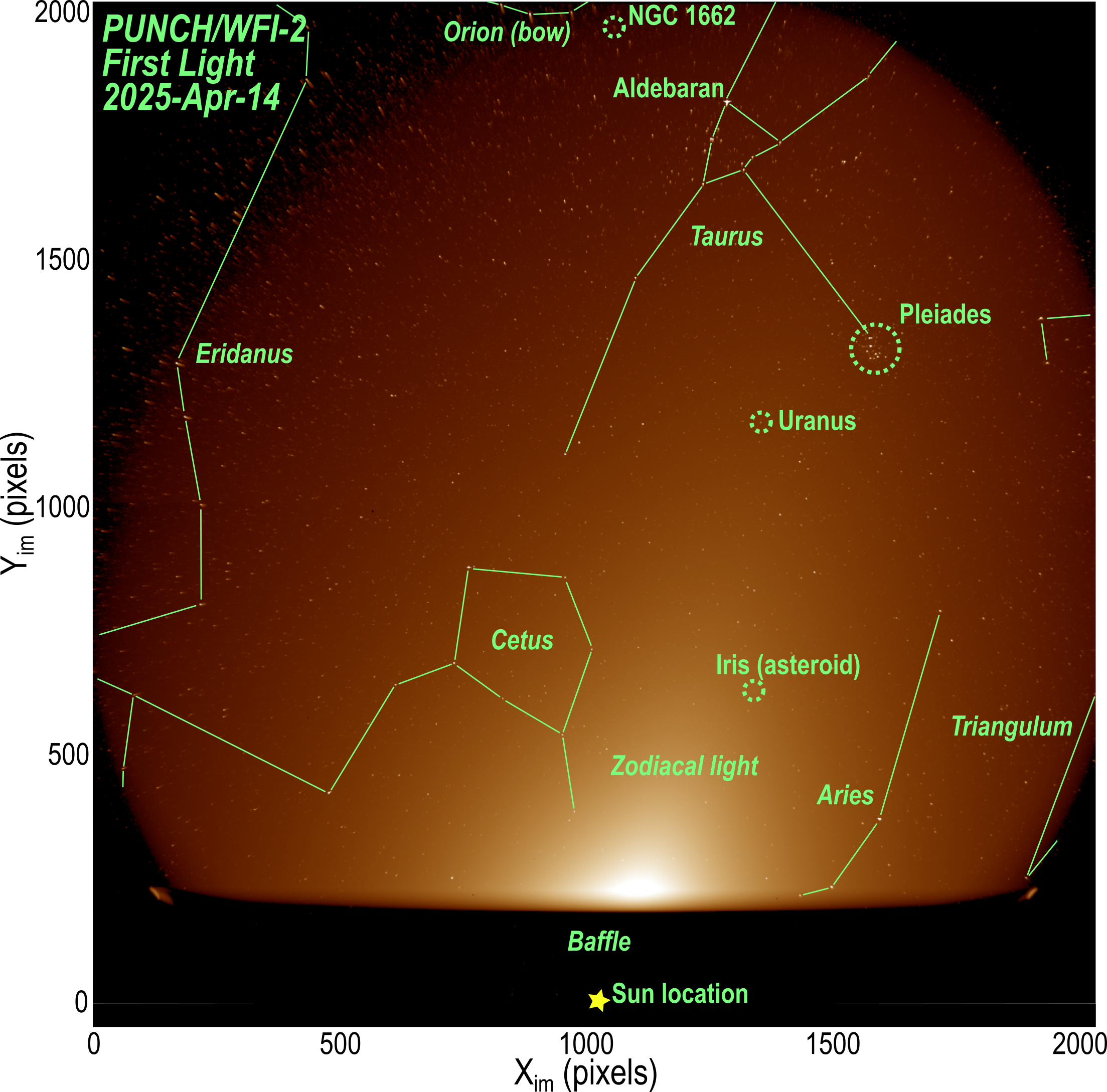}
    \caption{WFI-2 first light image from 14-April-2025 shows good focus, excellent stray light, and high sensitivity.  Constellations and the asteroid Iris (magnitude 9.5 on that date) are marked for reference.}
    \label{fig:wfi-first-light}
\end{figure}

The Wide Field Imagers passed testing and launched into space on board three PUNCH
spacecraft on 11-Mar-2025.  The instrument doors were opened one month later.  First
Figure \ref{fig:wfi-first-light} is the first-light image from WFI-2, showing good
focus, high sensitivity, and low stray light.  All three instruments showed coma in the 
extreme corners of the image, as expected; the coma is removed in ground processing at
the Science Operations Center \citep{hughes_etal_2025}.  

WFI performance was found to be very consistent across the three instruments. 
Exposure time is set to 51s for the standard science campaign. The instrument pointing
was adjusted on orbit so that the solar vanes are 3$^\circ$ to 3.5$^\circ$ from the Sun
(12-14 R$_\odot$) rather than the design separation of 4.5$^\circ$. 

The instrument design is, at root, a simple dioptric imaging system. The particular
benefits of WFI arise from the very deep baffle system, sensitive and 
reproducible-exposure CCD camera system, and ground processing.

WFI mosaics form the bulk of PUNCH science data, providing routine polarized images of
the solar wind over a 90$^\circ$ wide field of view centered on the Sun.

\begin{acks}[Acknowledgments]
PUNCH and WFI owe much of the overall design and scientific conept to early efforts by
T.A. Howard, whose work was instrumental in co-developing and exploiting methods
for deep background-subtraction of heliospheric images; in predicting and promoting
the possibility of wide-field 3-D analysis via polarization; and in helping demonstrate
the feasibility of heliospheric imaging from below the Van Allen belts.

We particularly acknowledge Tele Vue Optics and remember their founder, 
Al Nagler (1935-2025),
for designing and supplying the WFI focusing optics, working well 
outside Tele Vue’s “comfort zone” of eyepiece design, and stepping up to produce a 
very compact, achromatic, space-qualified system with very low distortion, several 
important stray-light reducing features, and excellent focus over a very wide field of view. 
PUNCH/WFI is but one of Al Nagler's many contributions to the astronomy, spaceflight, 
and scientific communities, and he is both sorely missed and fondly remembered.

We remember author RRG (1957-2024), who brought competence and good cheer to everyone who 
used his testing facilities at SwRI, and whose tireless effort helped bring WFI through its 
environmental testing campaign. Roy is and will continue to be missed.

We also gratefully acknowledge Chris Eyles and Russ Howard for early consultation on 
heliospheric imager design based on his experience with SMEI; and the SwRI Internal
Research
program for much early support that led to the successful PUNCH proposal. 

We thank the NFI team at the Naval Research Laboratory for their 
assistance and effort during SCOTCH testing of the EM WFI and WFI-1 and integration 
of the WFI PFW. 

PUNCH is a heliophysics mission to study the corona, solar wind, and space weather as 
an integrated system, and is part of NASA's Explorers program (Contract 80GSFC14C0014).

\end{acks}

\begin{ethics}
\begin{conflict}
The authors declare that they have no conflicts of interest.
\end{conflict}
\end{ethics}


\bibliographystyle{spr-mp-sola}
\bibliography{punch-bib}  

\end{document}